\documentclass[11pt]{article}

\usepackage{amsmath, amsthm, amssymb, graphicx}
\usepackage{booktabs, multicol, multirow} 
\usepackage[flushleft]{threeparttable} 
\usepackage{indentfirst} 
\usepackage{natbib}

\usepackage[hypertexnames=false]{hyperref}
\hypersetup{colorlinks, linkcolor = {teal}, citecolor = {blue}, urlcolor ={blue!80!black}}
\usepackage[dvipsnames,svgnames, x11names]{xcolor}

\usepackage{ulem}

\newcommand{\blind}{0}

\newtheorem{theorem}{Theorem}
\newtheorem{assumption}{Assumption}
\newtheorem{definition}{Definition}
\newtheorem{lemma}{Lemma}
\newtheorem{proposition}{Proposition}

\begin{document}
\def\spacingset#1{\renewcommand{\baselinestretch}%
{#1}\small\normalsize} \spacingset{1}


\if0\blind
{
  \title{\bf On the Size Control of the Hybrid Test for Predictive Ability}
  \author{Deborah Kim\hspace{.2cm}\\
    Department of Economics, Northwestern University, Evanston IL 60208\\ (deborahkim@u.northwestern.edu)
   }
  \maketitle
} \fi

\if1\blind
{
  \bigskip
  \bigskip
  \bigskip
  \begin{center}
    {\LARGE\bf On the Size Control of the Hybrid Test for Predictive Ability}
\end{center}
  \medskip
} \fi

\bigskip
\begin{abstract}
We analyze theoretical properties of the hybrid test for superior predictability. We demonstrate with a simple example that the test may not be pointwise asymptotically of level $\alpha$ at commonly used significance levels and may lead to rejection rates over $11\%$ when the significance level $\alpha$ is $5\%$. Generalizing this observation, we provide a formal result that pointwise asymptotic invalidity of the hybrid test persists in a setting under reasonable conditions. As an easy alternative, we propose a modified hybrid test based on the generalized moment selection method and show that the modified test enjoys pointwise asymptotic validity. Monte Carlo simulations support the theoretical findings.
\end{abstract}

\noindent%

\spacingset{1.45} 
\section*{{1. Introduction}}\label{sec:intro}
A test of superior predictive ability (SPA) compares many forecasting methods. More precisely, it tests whether a certain forecasting method outperforms a finite set of alternative forecasting methods. Most notably in the literature of tests of SPA, \cite{white2000} developed a framework for a SPA test and proposed a SPA test called the Reality Check for data snooping. \cite{hansen2005test} proposed a SPA test featuring improved power in the framework of \cite{white2000}. Finally, \cite{song2012testing} devised a SPA test, called the hybrid test, which delivers better power against certain local alternative hypotheses under which both of the SPA tests of \cite{white2000} and \cite{hansen2005test} perform poorly.

One of the main challenges of all SPA tests lies in finding a suitable critical value. This is because popular test statistics used in this setting are asymptotically non-pivotal and depend on parameters that cannot be consistently estimated. More concretely, the null hypothesis of a SPA test can be written as $H_0: \mu \leq 0_M$ for a parameter $\mu \in \mathbb{R}^M$ where $0_M$ is a $M$-dimensional vector with zeros, inequality applies elementwise, and $M\geq 2$ in general. It then follows that the limiting distribution of standard test statistics depends on exactly which of the elements in the vector $\mu$ are equal to zero. This property prevents researchers from using any tabulated critical values. 

To circumvent the above problem, \cite{white2000} proposed to use a critical value from the so-called least favorable distribution in his Reality Check test. The approach exploits the fact that the distribution of its test statistic $T$ under $\mu=0_M$ is stochastically largest over all possible null distributions satisfying $\mu\leq 0_M$. The distribution under $\mu=0_M$ is then called the least favorable case. The Reality Check approximates the least favorable distribution using the bootstrap and takes the $1-\alpha$ quantile of the distribution as the critical value, where $\alpha$ is a significance level. The resulting critical value converges to a value which is always larger than or equal to $1-\alpha$ quantile of the limiting distribution of $T$ under any null distribution and thus the approach yields a test with correct asymptotic size. 

\cite{song2012testing} followed the Reality Check in the construction of the hybrid test and used the same least favorable distribution, i.e., the one associated with $\mu=0_M$. However, in this article, we show that this null distribution is not the least favorable one for the type of test statistic of the hybrid test which, in particular, combines two different test statistics. Whereas one of the test statistics is stochastically largest under $\mu=0_M$, the other one is not. Consequently, the hybrid test which employs bootstrap approximations to the distribution with $\mu=0_M$ fails to control the rejection probability under the null and loses pointwise asymptotic validity. 

As the main contribution of this article, we show that the hybrid test may not be pointwise asymptotically of level $\alpha$ under reasonable conditions. This implies that a researcher could reject the null hypothesis $H_0$ with probability higher than the significance level $\alpha$ in the limit even when $H_0$ is true. A simple example in \hyperref[subsec:example]{Section 3.1} shows that the rejection probability under the null distribution could be over $11\%$ when the significance level $\alpha$ is set at $5\%$. Our results illustrate that the cause of the problem lies behind the fact that the bootstrap procedure that the hybrid test uses approximates neither the asymptotic distribution of the test statistics nor their least favorable distribution. As an easy fix, we propose a modified hybrid test which is pointwise asymptotically of level $\alpha$, again under reasonable conditions. Our proposed modification follows the generalized moment selection method by \cite{andrews2010inference} after accounting for the fact one of the test statistics in the hybrid test does not exhibit certain monotonicity properties that are required for the generalized moment selection method.

Neither devising an alternative test to the hybrid test nor analyzing theoretical properties of the modified hybrid test is the main focus of this article. Yet, it is worth making a remark on asymptotic validity in SPA tests. When it comes to controlling the asymptotic type I error, uniform asymptotic validity, which is a necessary condition of pointwise asymptotic validity, is regarded as the gold-standard as it guarantees uniform control of Type I error over all null distributions at a fixed sample size. For such a reason, the literature of testing moment inequalities has addressed the importance of uniform asymptotic validity. However, the property has not been yet discussed in depth in the literature of SPA tests, of which the reason we attribute to complexity coming from dealing with dependent data.\footnote{Not only the papers discussed in this article, \cite{white2000} and \cite{hansen2005test}, but also a recent paper on conditional SPA test by \cite{LiEtAl2021REStud} deals with pointwise asymptotic validity, not uniform asymptotic validity. For readers who are interested in uniform asymptotic validity in the context of testing moment inequality, see \cite{canayshaikh}.} As a first step of engaging uniform asymptotic validity in this literature, we formally define uniform asymptotic validity in the context of SPA tests following \cite{andrews2010inference} and show that the modified hybrid test is not uniformly asymptotically valid in \hyperref[sec:appendixC]{Appendix C}. Developing a general framework for uniform asymptotic validity for SPA tests could be interesting, yet we leave it as a topic for future research.

This article is organized as follows. \hyperref[sec:setup]{Section 2} lays out notation and describes the hybrid test as originally proposed by Song (2012). \hyperref[sec:body]{Section 3} presents the main result of the properties of the hybrid test and proposes the modified hybrid test with its formal properties. \hyperref[sec:montecarlo]{Section 4} explores the Monte Carlo simulations of the hybrid test and the modified one. Lastly, \hyperref[sec:conclusion]{Section 5} concludes. The proofs of the formal results are included in \hyperref[sec:appendixB]{Appendix B}. 

\section*{{2. The hybrid test for predictive ability}}\label{sec:setup}
In this section, we introduce the hybrid test proposed by \cite{song2012testing} in a simple setup as in \cite{hansen2005test} where the data generating process is stationary. The framework in \cite{song2012testing} incorporates more general settings, yet a simple setup better serves our purpose by allowing us to solely focus on verification of the theoretical property of the hybrid test. We explain how our setup is different from the original one in more detail in \hyperref[sec:body]{Section 3}.

Consider a situation where we aim to predict a $\tau$-ahead unknown random variable $\xi_{t+\tau}$ at time $t$. Suppose that we have $M+1$ different forecasts: a benchmark $\varphi_{t,0}$ and a finite set of alternative forecasts $\varphi_{t,m}$, $m\in\mathbf{M} \equiv \{1, \ldots, M\}$. The objective of the hybrid test is to test whether the benchmark forecast is superior to all other alternative forecasts in terms of predictive ability. To compare the predictive ability, we assess the risk (of prediction) of the $m$th forecast in terms of expected risk, $E[\Lambda(\xi_{t+\tau}, \varphi_{t,m})]$ for $m=0, 1, \ldots, M$. An example of such risk is the squared error $\Lambda(\xi_{t+\tau}, \varphi_{t,m})= \|\xi_{t+\tau}-\varphi_{t,m}\|^2$ for $ \xi_{t+\tau}, \varphi_{t,m}\in \mathbb{R}$. For more examples, refer to \cite{song2012testing} or references therein. 

If the expected risk of the forecast $\varphi_{t,m}$ is greater than or equal to that of the benchmark $\varphi_{t,0}$, we say the benchmark forecast $\varphi_{t,0}$ dominates the forecast $\varphi_{t,m}$ (in terms of predictive ability). Let us define the relative risk variable between the benchmark forecast $\varphi_{t,0}$ and the $m$th alternative forecast $\varphi_{t,m}$ as
\begin{align*}
    d_{t,m} = \Lambda(\xi_{t+\tau}, \varphi_{t,0})- \Lambda(\xi_{t+\tau}, \varphi_{t,m}) \text{ for }m=1, \ldots,M.
\end{align*}
Then the null hypothesis that the benchmark forecast $\varphi_{t,0}$ dominates all alternative forecasts in $\mathbf{M}$ can be formulated as
\begin{align}
H_0: E[d_{t,m}] \leq 0 \text{ for all } m \in \mathbf{M} \text{ and }
H_1: E[d_{t,m}] > 0 \text{ for some } m \in \mathbf{M}.\label{eq:hypotheses}
\end{align}
The relative risk variables are building blocks for our analysis. We regard $\{d_t: t= 1, \ldots, n\}$ where $d_t =(d_{t,1}, \ldots, d_{t,M})^t \in \mathbb{R}^M$ as our observations and abstract away from how they are constructed. We assume $\{d_t: t= 1, \ldots, n\}$ are stationary under distribution $F$ with mean $\mu\equiv E[d_t]$ and thus expectation in \eqref{eq:hypotheses} is taken with respect to $F.$

Before proceeding, it is useful to define what we mean by pointwise asymptotically valid tests. A test $\phi_n=\phi_n(d_1, \ldots, d_n)$ for the null hypothesis $H_0$ is said to be pointwise asymptotically of level $\alpha$ if it satisfies
\begin{align}
\limsup _{n \rightarrow \infty} E\left[\phi_n\right] \leq \alpha \label{eq:ptlevel}
\end{align}
for a data generating process satisfying the null hypothesis $H_0$ and the maintained assumptions. Interchangeably, we say $\phi_n$ attains pointwise asymptotic validity. If a test fails to satisfy condition \eqref{eq:ptlevel}, then we can always find some distribution $F$ under $H_0$ along which the rejection probability $E[\phi_n]$ exceeds the significance level $\alpha$ infinitely often as the sample size $n$ grows. 


An unusual feature of the hybrid test is that the test uses two pairs of a test statistic and a critical value. In \hyperref[sec:body]{Section 3}, we show how this feature complicates theoretical analysis of pointwise asymptotic validity. The first pair $(\hat{T}^r_n, \hat{c}^{r*}_n)$ is adopted from the Reality Check which tests the one-sided null hypothesis \eqref{eq:hypotheses}. We call $\hat{T}^r_n$ and $\hat{c}^{r*}_n$ one-sided test statistic and one-sided critical value, respectively. The second pair $(\hat{T}^s_n, \hat{c}^{s*}_n)$ is adopted from the symmetrized test by \cite{lmw2005}. We call $\hat{T}^s_n$ and $\hat{c}^{s*}_n$ two-sided test statistic and two-sided critical value. The name, two-sided test, comes from the fact that the two-sided statistic $\hat{T}^s_n$ is originally proposed to test a two-sided null hypothesis $H^s_0:  \mu \leq 0_M \text{ or } \mu \geq 0_M$.

The test statistics and critical values are defined through two functions, $T^r: \mathbb{R}^M \mapsto  \mathbb{R}$ and  $T^s: \mathbb{R}^M \mapsto  \mathbb{R}$ defined as 
\begin{align}
    T^r(x) \equiv \max_{m \in \mathbf{M}} x_m \quad\text{ and }\quad
     T^s(x) \equiv \text{min}(\max_{m \in \mathbf{M}} x_m, \max_{m \in \mathbf{M}} -x_m), \label{eq:statfun}
\end{align}
and through two statistics 
\begin{align*}
    \hat{d}_n = \frac{1}{n}\sum^n_{t=1} d_t \quad\text{ and }\quad\hat{D}_n = \text{diag}(\hat{\sigma}^2_{n,1}, \ldots, \hat{\sigma}^2_{n,M})
\end{align*}
where $\hat{\sigma}^2_{n,m}, m=1, \ldots,M$ are some estimators for the asymptotic variance of $\sqrt{n}(\hat{d}_n - \mu).$ For now, we assume that these statistics satisfy
\begin{align}
    \sqrt{n} (\hat{d}_n - \mu) \overset{d}{\to} N(0_M, \Sigma) \quad\text{ and }\quad \hat{\sigma}^2_{n,m} \overset{p}{\to} \Sigma_{m,m}
    \label{eq:AsyNormality}
\end{align}
for all $m=1,\ldots,M$ where $\Sigma$ is a $M\times M$ semi-positive matrix and $\Sigma_{i,j}$ is the $(i,j)$ component of $\Sigma$. Note $\hat{D}_n$ is an estimator of $\text{diag}(\Sigma)$. The two test statistics of the hybrid test are
\begin{align*}
&\hat{T}^q_n \equiv T^q\left(\sqrt{n}\hat{D}^{-1/2}_{n}\hat{d}_n\right)   \text{ for }q \in \{r,s\}.
\end{align*}

Defining critical values is not as straightforward as defining the test statistics. Define two values $\bar{c}^r(\alpha, \gamma)$ and $\bar{c}^s(\alpha, \gamma)$ satisfying these two conditions:
\begin{align*}
&\lim_{n \to \infty} P\left\{\hat{T}^r_n > \bar{c}^r(\alpha, \gamma) \text{ and }\hat{T}^s_n \leq \bar{c}^s(\alpha, \gamma)\right\} = \alpha(1-\gamma) \text{ and }\\
&\lim_{n \to \infty} P\left\{\hat{T}^s_n > \bar{c}^s(\alpha, \gamma)\right\}=\alpha\gamma
\end{align*}
for a significance level $\alpha \in (0,1)$ and a fixed tuning parameter $\gamma \in (0,1)$. If the critical values $\hat{c}^{r*}_n $ and $\hat{c}^{s*}_n$ were consistent for $\bar{c}^r(\alpha, \gamma)$ and $\bar{c}^s(\alpha, \gamma)$, the hybrid test would be asymptotically pointwise of level $\alpha$. The two values are, however, tricky to estimate. To see this, let us rewrite the test statistics as
\begin{align*}
    \hat{T}^q_n \equiv T^q\left(\sqrt{n}\hat{D}^{-1/2}_{n}(\hat{d}_n-\mu) + \sqrt{n}\hat{D}^{-1/2}_{n}\mu\right)   \text{ for }q \in \{r,s\}.
\end{align*}
Because $T^q$ is a continuous mapping for any $q\{r,s\}$, the weak convergence of $\hat{T}^q_n$ follows from the weak convergence of the argument. The first term in the argument $\sqrt{n}\hat{D}^{-1/2}_{n}(\hat{d}_n-\mu)$ is stochastically bounded by \eqref{eq:AsyNormality}, yet we cannot consistently estimate the second term $\sqrt{n}\hat{D}^{-1/2}_{n}\mu$ unless $\mu=0_M$ because at least one component of $\mu$ diverges. 

Instead of directly dealing with the tricky part, the Reality Check takes an indirect approach based on the least favorable case to define a critical value. To explain, it is useful to define a distribution function
\begin{align}
    J^q_n(x; \eta) = P\left\{T^q (\hat{D}^{-1/2}_n \sqrt{n} (\hat{d}_n -\mu) + \hat{D}^{-1/2}_n\sqrt{n}\eta) \leq x \right\}\label{eq:def_distfn}
\end{align}
for any $x, \eta \in \mathbb{R} $ for $q \in \{r,s\}$. Note $J^q_n(x;\mu)$ is simply the distribution function of the test statistic $\hat{T}^q_n$ for $q \in \{r,s\}$. The Reality Check then takes advantage of the fact that $T^r(\hat{D}^{-1/2}_n \sqrt{n} (\hat{d}_n - \mu))$ can be consistently estimated by conventional data-dependent methods and it stochastically dominates the one-sided test statistic $\hat{T}^r_n$ (at first order) under $H_0$, i.e., 
\begin{align*}
    J^r_n (x; 0_M) \leq J^r_n (x;\mu) \quad \text{ for all } x\in \mathbb{R} \text{ and for any }\mu \leq 0_M,
\end{align*}
which immediately follows from monotonicity of $T^r(\cdot).$ Equivalently, any quantile of $T^r(\hat{D}^{-1/2}_n$ $\sqrt{n} (\hat{d}_n - \mu))$ is greater than or equal to that of $\hat{T}^r_n$ under $H_0$, i.e.,
\begin{align}
    (J^r_n)^{-1} (t; 0_M) \geq  (J^r_n)^{-1} (t;\mu) \quad \text{ for all }t \in [0,1] \text{ and for any }\mu \leq 0_M \label{eq:lfc}
\end{align}
where $ (J^q_n)^{-1} (t; \eta) = \inf\{x \in \mathbb{R}: J^q_n(x;\eta) \geq 1-t\}$ for any $t \in [0,1]$ and for $q\in \{r,s\}$. In this sense, the distribution of $T^r(\hat{D}^{-1/2}_n \sqrt{n} (\hat{d}_n - \mu))$, or the distribution of $\hat{T}^r_n$ at $\mu=0_M$ is called \textit{the least favorable case} of $\hat{T}^r_n$ under $H_0$. The Reality Check defines its critical value $\underbar{c}^r$ as the $1-\alpha$ quantile of the limit distribution of the least favorable case, i.e., $\lim_{n\to \infty} J^r_n (\cdot; 0_M) $. As intended, the stochastic dominance then guarantees the test $1\{ \hat{T}^r_n >\underbar{c}^r \}$ being asymptotically of level $\alpha$.

Following the Reality Check, the hybrid test defines the bootstrap-based critical value $\hat{c}^{q*}_n$ as a consistent estimator of the $1-\alpha$ quantile of $\lim_{n\to \infty} J^r_n (\cdot; 0_M) $ for $q \in \{r,s\}$. Here we explain the procedure step by step.
Consider a bootstrap sample $\{\hat{d}_{n,b}^{*}:  1\leq b \leq B\}$ obtained by the stationary bootstrap by \cite{RomanoPolitis1994JASA} where we denote the $m$th element of $\hat{d}_{n,b}^{*}$ as $\hat{d}_{n,b,m}^{*}$. Define a centred bootstrap sample as
\begin{align}
\left\{\tilde{d}^*_{n,b} \equiv \hat{d}_{n,b}^{*}-\hat{d}_n: b= 1, 2, \ldots, B\right\}.
\label{eq:centredmean}
\end{align}
Define the bootstrap test statistics as $\{(\hat{T}_{n,b}^{r *}, \hat{T}_{n,b}^{s *})\}_{b=1}^{B}$ where $\hat{T}^{q*}_{n,b}$ is
\begin{align}
&\hat{T}^{q*}_{n,b}
\equiv T^q \left(\sqrt{n} \hat{D}^{-1/2}_n \tilde{d}^{*}_{n,b} \right) \quad \text{ for }q \in \{r,s\}\label{eq:bootstat}
\end{align}
and $\{\hat{\sigma}_{n,m}: m \in \mathbf{M} \}$ are not bootstrapped. Choose $\gamma \in (0,1]$. The hybrid test defines the two-sided critical value $\hat{c}^{s*}_n$ as the $1-\alpha\gamma$ quantile of the bootstrap sample $\{\hat{T}^{s*}_{n,b}\}^B_{b=1}$, i.e.
\begin{align*}
\hat{c}^{s*}_n\equiv 
\inf \left\{x \in \mathbb{R}: \frac{1}{B}\sum^B_{b=1} 1\left\{\hat{T}^{s*}_{n,b} \leq x\right\}\geq 1-\alpha \gamma \right \}.
\end{align*} 
Given $\hat{c}^{s*}_n $, the one-sided critical value $\hat{c}^{r*}_n $ is defined as the $1-\alpha(1-\gamma)$ quantile of the bootstrap sample $\{\hat{T}^{r*}_{n,b}  \cdot 1\{\hat{T}^{s*}_{n,b} \leq \hat{c}^{s*}_{n} \}\}^B_{b=1}$ for $\gamma \in (0,1)$, i.e.
\begin{align*}
\hat{c}^{r*}_{n}  \equiv 
\inf\left\{ x \in \mathbb{R}: \frac{1}{B}\sum^B_{b=1} 1\left\{ \hat{T}^{r*}_{n,b}  \cdot 1\left\{\hat{T}^{s*}_{n,b} \leq \hat{c}^{s*}_{n} \right\}
 \leq x \right\}\geq 1-\alpha (1-\gamma) \right \},
\end{align*}
and $\hat{c}^{r*}_n= \infty$ for $\gamma=1.$

Finally, given the two pairs, $(T^r_n, \hat{c}^{r*}_n)$ and $(T^s_n, \hat{c}^{s*}_n)$, the hybrid test rejects the null hypothesis if $T^r_n>\hat{c}^{r*}_n$ or $T^s_n> \hat{c}^{s*}_n$. For brevity, define $\phi^r_n \equiv 1\{T^r_n>\hat{c}^{r*}_n\}$ and $\phi^s_n\equiv 1\{T^s_n> \hat{c}^{s*}_n\}$, say one-sided test and two-sided test. Then the hybrid test is defined as 
\begin{align}\label{eq:hybridtest}
\phi_n \equiv 1\{\hat{T}^r_n >  \hat{c}^{r*}_n \text{ or }\hat{T}^s_n >  \hat{c}^{s*}_n\}= \phi^r_n (1-\phi^s_n ) +\phi^s_n.
\end{align}
That is, the rejection of the hybrid test is the union of the two rejection regions by $\phi^r_n$ and $\phi^s_n.$ The tuning parameter $\gamma \in (0,1]$ determines how much the two-sided test $\phi^s_n$ contributes to forming the rejection region as opposed to the one-sided test $\phi^r_n.$ If $\gamma $ is zero, the hybrid test coincides with the one-sided test $\phi^r_n$. If $\gamma $ is 1, then the hybrid test corresponds to the two-sided test $\phi^s_n$. We restrict the tuning parameter $\gamma$ to be in $(0,1]$ as the asymptotic properties of the one-sided test can be found in \cite{white2000}.

\section*{{3. On the asymptotic validity of the hybrid test}}\label{sec:body}
In this section, we investigate asymptotic properties of the hybrid test under the null hypothesis, which are not formally discussed in \cite{song2012testing}. First, we provide a simple example where the rejection probability of the hybrid test exceeds significance level $\alpha$ in the limit. Then we present the main results generalizing the observation made in the example. 

The following set of assumptions strengthens the setup of \cite{song2012testing} and facilitates our theoretical analysis of the hybrid test.
\begin{assumption}
$\{d_t\in \mathbb{R}^M: t\leq n\}$ is stationary under distribution $F$ with $E[d_t]< \infty$. \label{as:stationarity}
\end{assumption}
\begin{assumption} $\sqrt{n}(\hat{d}_n-\mu) \overset{d}{\to} N(0_M, \Sigma)$ as $n\to\infty$.
\label{as:weakconv}
\begin{assumption} The bootstrap sample in \eqref{eq:centredmean} satisfies
\begin{align*}
\sup_{z \in \mathbb{R}^{M}}\left|P^{*}_n\left\{\sqrt{n}\left(\hat{d}^*_{n,b}-\hat{d}_n\right) \leq z\right\}
-P\left\{\sqrt{n}(\hat{d}_n-\mu) \leq z\right\}\right| \stackrel{p}{\rightarrow} 0 \text{ as }n \to \infty
\end{align*}
where $P^*_n$ denotes the probability measure conditional on the sample $\{d_t\}^{n}_{t=1}$. \label{as:bconsistency}
\end{assumption}
\end{assumption}
\begin{assumption}
$\Sigma_{m,m}>0$ for $m=1, \ldots, M$. \label{as:psd_cov}
\end{assumption}
\begin{assumption} $\hat{\sigma}^2_{n,m} \overset{p}{\to} \Sigma_{m,m}$ as $n\to \infty$ for $m=1, \ldots, M$.
\label{as:consist_var}
\end{assumption}
Assumption \ref{as:stationarity} implies that the marginal distribution of $d_t$ does not vary over time and has a finite mean. This ensures that null hypothesis $H_0$ in \eqref{eq:hypotheses} is well-defined. Stationarity of data generating processes is often made in the literature so as to invoke asymptotic normality and bootstrap consistency (for example, Assumption 1 in \cite{hansen2005test} and Assumption A in \cite{white2000}). The asymptotic normality in Assumption \ref{as:weakconv} is necessary in order to enable the inference of the test statistics. \cite{song2012testing} requires asymptotic normality of a generic statistic $\tilde{d}_n$ in place of $\hat{d}_n$, yet we consider the case where $\tilde{d}_n$ is given as a sample mean $\hat{d}_n$ as in \cite{white2000} and \cite{hansen2005test}. The bootstrap consistency in Assumption \ref{as:bconsistency} guarantees that the bootstrap approximates the distribution of $\sqrt{n}(\hat{d}_n-\mu)$ for large $n$.
This is necessary to justify that the hybrid test uses the bootstrap critical values. Assumption \ref{as:weakconv} and \ref{as:bconsistency} can be attained by imposing $\alpha$-mixing condition onto the process $\{d_t\}^n_{t=1}$ as in \cite{hansen2005test}, yet we maintain high-level assumptions building on the literature. Assumption \ref{as:psd_cov} guarantees that no element of $\sqrt{n}(\hat{d}_n-\mu)$ degenerates in the limit, which is made to innocuously simplify the analysis. Finally, another high-level Assumption \ref{as:consist_var} assumes consistency of the variance estimator, which is implicitly assumed in \cite{song2012testing} in his use of studentized statistics.

\color{black}\subsection*{3.1 An Example}\label{subsec:example}
To gain intuitions on the asymptotic properties of the hybrid test, we consider a simple example where the number of alternative forecasts is two, $M=2$. Consider the distribution function $F$ in Assumption \ref{as:stationarity} such that $\mu_{1}=0$ and $\mu_{2}<0$. Namely, the first alternative forecast is as risky as the benchmark, whereas the benchmark dominates the second alternative forecast. We further assume that the covariance matrix in Assumption \ref{as:weakconv} is the identify matrix, $\Sigma=I_2$, and it is known. Thus we simply use $\hat{D}_n = I_2$. For simplicity, let $\gamma=0.5$.

First, we derive the asymptotic distribution of the test statistics, $\hat{T}^{r}_{n}$ and $\hat{T}^{s}_{n}$. By Assumption \ref{as:weakconv}, we have
\begin{align*}
\sqrt{n}( 
\hat{d}_{n,1} - \mu_{1},
\hat{d}_{n,2}- \mu_{2}
) \overset{d}{\to} Z\equiv \left(Z_1, Z_2\right) \sim N(0, I_2).
\end{align*}
This and the condition that $\mu_{1}=0, \mu_{2}<0$ together imply that $\sqrt{n}\hat{d}_{n,2}$ diverges to $-\infty$ as $n\to\infty$ while $\sqrt{n}\hat{d}_{n,1}$ is stochastically bounded. Then the two test statistics depend only on $\sqrt{n}\hat{d}_{n,1}$ for large $n$, eventually yielding the following approximation:
\begin{align}
\left( 
\begin{matrix}
\hat{T}^{r}_{n} \\ \hat{T}^{s}_{n}
\end{matrix}
\right)
\equiv
 \left(
\begin{matrix}
\sqrt{n}\max(\hat{d}_{n,1}, \hat{d}_{n,2}) \\
\sqrt{n}\min(\max(\hat{d}_{n,1}, \hat{d}_{n,2}), \max(-\hat{d}_{n,1}, -\hat{d}_{n,2}) )
\end{matrix}
\right)
\approx  \left(  
\begin{matrix}
\sqrt{n}\hat{d}_{n,1} \\ \sqrt{n}\hat{d}_{n,1}
\end{matrix}
\right)
\overset{d}{\to} \left(  
\begin{matrix}
Z_1\\Z_1
\end{matrix}
\right). \label{eq:exampletstat}
\end{align}

Meanwhile, $(\hat{T}^{r*}_{n,b}, \hat{T}^{s*}_{n,b})$ weakly converges to a distribution different from $(Z_1, Z_1)$ in \eqref{eq:exampletstat}. According to Assumption \ref{as:bconsistency}, we have 
\begin{align*}
\sqrt{n}( 
\tilde{d}^*_{n,b,1},
\tilde{d}^*_{n,b,2}
) \overset{d}{\to} \left(V_1, V_2\right) \sim N(0, I_2)
\end{align*}
with probability approaching 1, and then continuous mapping theorem gives
\begin{align*}
(\hat{T}^{r*}_{n,b},
\hat{T}^{s*}_{n,b} )
\equiv (
T^r (\sqrt{n} \tilde{d}^*_{n,b}),
T^s (\sqrt{n} \tilde{d}^*_{n,b})
)
\overset{d}{\to} (T^r(V),T^s(V))
\end{align*}
with probability approaching 1 where $V\equiv (V_1, V_2) \sim N(0_2, I_2)$. 

Our salient finding comes from a comparison between the two asymptotic distributions of $\hat{T}^s_n$ and $\hat{T}^{s*}_{n,b}$: the marginal asymptotic distribution of $\hat{T}^{s*}_{n,b}$ does not stochastically dominate that of $\hat{T}^s_n$. In fact, the two distribution functions of $Z_1$ and $T^s(V)$, which are the limit distributions of $\hat{T}^s_n$ and $\hat{T}^{s*}_{n,b}$, cross each other. This results in
\begin{align}
  \Phi^{-1}(t)> (J^s)^{-1} (t)  \text{ for some }t \in (0,1)\label{eq:violation_lfc_ex}
\end{align}
where $\Phi$ and $J^s$ are the distribution functions of $Z_1$ and $T^s(V)$. This rather unexpected result suggests that the distribution of $\hat{T}^s_n$ at $\mu=0_2$ that the bootstrap test statistic $\hat{T}^{s*}_{n,b}$ approximates is not the least favorable case of $\hat{T}^s_n$ under the null hypothesis $\mu \leq 0_2$ for large $n$. This contrasts to the fact that the distribution $\hat{T}^r_n$ at $\mu=0_2$ is the least favorable case of $\hat{T}^r_n$ under $\mu \leq 0_2$ for any $n$ in the sense of \eqref{eq:lfc}. 

The finding implies that the hybrid test may not attain pointwise asymptotic validity. Simple algebra provides a formula for the distribution function: $J^s(x) =-2\Phi^2(x) + 4\Phi(x) -1$ for $x \in [0, \infty)$. It is easy to show that the two-sided critical value $\hat{c}^{s*}_n$ is consistent for the $1-\alpha/2$ quantile of $J^s$, 
i.e. 
\begin{align}
\hat{c}^{s*}_n \overset{p}{\to} c^s (\alpha/2)\equiv \inf\{x \in \mathbb{R}:  J^s(x) \geq 1-\alpha/2\}. \label{eq:examplecv}
\end{align}
Then \eqref{eq:exampletstat} and \eqref{eq:examplecv} combine to give
\begin{align*}
E[\phi_n] 
&= P\{ \hat{T}^r_n > \hat{c}^{r*}_n \text{ or } \hat{T}^s_n > \hat{c}^{s*}_n \} \approx P\{Z_1 > \min(c^r (\alpha/2), c^s (\alpha/2))\} \geq P\{Z_1 > c^s (\alpha/2)\} 
\end{align*}
where $c^r(\alpha/2)$ is the probability limit of $\hat{c}^{r*}_n$ and the inequality holds by the definition of minimum. The last term $1-\Phi(c^s(\alpha/2))$ exceeds the significance level $\alpha$ if $\alpha\in (0, 0.25)$. The gap between $E[\phi_n]$ and $\alpha$, the violation of pointwise asymptotic validity of the hybrid test, could be substantive: $1-\Phi(c^s(\alpha/2))$ is 0.158, 0.112, and 0.05 for different values of $\alpha$, 0.10, 0.05, and 0.01 respectively.

\subsection*{3.2. Main Results}\label{subsec:mainresult}
The example in the previous subsection shows that, given all the assumptions are maintained, there exist distributions $F$ under which the distribution of $\hat{T}^s_n$ at $\mu=0_M$ is not the least favorable case of $\hat{T}^s_n$ under the null hypothesis in \eqref{eq:hypotheses} for large $n$ in the sense of \eqref{eq:lfc}. This phenomenon, in fact, holds in general and it leads to the hybrid test not being pointwise asymptotically valid. We start with introducing Lemma \ref{lemma:LFC}, of which the proof is delegated to \hyperref[sec:appendixB]{Appendix B}.

\begin{lemma}
\label{lemma:LFC}
Consider $T^s: \mathbb{R}^M \mapsto \mathbb{R}$ defined in \eqref{eq:statfun}. Suppose Assumption \ref{as:stationarity}, \ref{as:weakconv}, \ref{as:psd_cov} and \ref{as:consist_var} hold. Then there exists a distribution $F$ satisfying the null hypothesis in \eqref{eq:hypotheses} under which the limiting distribution of $T^s(\hat{D}^{-1/2}_n\sqrt{n}(\hat{d}_n-\mu))$ does not stochastically dominate that of the two-sided test statistic $\hat{T}^s_{n} \equiv T^s(\hat{D}^{-1/2}_n \sqrt{n} \hat{d}_n)$. 
\end{lemma}
Generalizing the result in \eqref{eq:violation_lfc_ex}, this lemma says that under the distribution $F$ and for sufficiently large $n$
\begin{align*}
    (J^s_n)^{-1}(t;0_M) < (J^s_n)^{-1}(t;\mu)  \text{ for some }t\in [0,1] \text{ for some }\mu \leq 0_M
\end{align*}
where $J^s_n$ is defined in \eqref{eq:def_distfn}. In other words, the distribution of $\hat{T}^s_n$ at $\mu=0_M$ is not the least favorable case of $\hat{T}^s_n$ under $\mu \leq 0_M$ for large $n$ in the sense of \eqref{eq:lfc}.

The implication that immediately follows from Lemma \ref{lemma:LFC} is that the two-sided test $\phi^s_n$ alone fails to control the size. If the tuning parameter $\gamma \in (0,1]$ is $1$, then the hybrid test coincides with the two-sided test, i.e $\phi_n= \phi^s_n$, and so the hybrid test is not pointwise asymptotically of level $\alpha.$ Rather unexpectedly, the following theorem tells us that the over-rejection of the null hypothesis driven by the two-sided test is overriding even for small $\gamma$ and consequently the hybrid test is not pointwise asymptotically valid for any $\gamma \in (0,1]$ for some $\alpha$. 

Before proceeding, we define the parameter space $\mathcal{F}$ as
\begin{align*}
    \mathcal{F} &\equiv \{(\Sigma, F): \text{ Assumption \ref{as:stationarity}, \ref{as:weakconv}, \ref{as:bconsistency}, \ref{as:psd_cov} and \ref{as:consist_var} are satisfied.}\}
\end{align*}
and define the subset of $\mathcal{F}$ which satisfies the null hypothesis as $\mathcal{F}_0 \equiv \{(\Sigma, F) \in \mathcal{F}: E[d_t] \leq 0_M\}$.

\begin{theorem}
Suppose that Assumption \ref{as:stationarity}, \ref{as:weakconv}, \ref{as:bconsistency}, \ref{as:psd_cov} and \ref{as:consist_var} hold. Let $2\leq M<\infty$. Suppose $(\Sigma, F) \in \mathcal{F}_0$ satisfies the following conditions:
\begin{enumerate}
\item there exists $m \in \mathbf{M}$ such that $\mu_m=0$, 
\item there exists $m' \in \mathbf{M}$ such that $\mu_{m'}<0$, and
\item $\Sigma$ is a diagonal matrix. 
\end{enumerate}
For any $\gamma \in (0,1]$, there exists an upper bound $\bar{\alpha} \equiv \bar{\alpha}(F,\gamma) \in (0, 0.5]$ such that the following condition holds:
        \begin{align*}
           \lim_{n \to \infty} E[\phi_n] > \alpha \text{ for any }\alpha \in (0, \bar{\alpha})
        \end{align*}        \label{prop:mains}
where $\phi_n$ is the hybrid test defined in (\ref{eq:hybridtest}). 
\end{theorem}  
Theorem \ref{prop:mains} claims that the hybrid test is not pointwise asymptotically of level $\alpha$ for any $\alpha \in (0, \bar{\alpha})$ for some $\bar{\alpha}$ under the parameterization $\mathcal{F}$. More specifically, it provides three sufficient conditions of a data generating process $(\Sigma, F)$ under which the asymptotic rejection probability exceeds the significance level $\alpha \in (0,\bar{\alpha})$. This has a substantial implication in practice: one may reject the null hypothesis at a higher rate than the desired level $\alpha$ even for large $n$ when the null hypothesis holds true. This result further implies that any power gain that the hybrid test is reported to possess over other SPA tests may be due to over-rejection of the hybrid test.

The magnitude of the upper bound $\bar{\alpha}$ could be of practical interest as one can carry out the hybrid test without taking the risk of committing the type I error over the conventional significance levels if $\bar{\alpha}$ is smaller than 0.01. The value of $\bar{\alpha}$ is, however, a priori unknown as it relies on $M$ as well as the number of the alternative forecasts which attain the same risk as the benchmark, i.e. $M_0 \equiv |\{m \in \mathbf{M}: \mu_m=0\}|$. In fact, once $M$, $M_0$ and $\gamma$ are fixed, $\bar{\alpha}$ can be obtained by numerical approximation. We tabulated some values of $\bar{\alpha}$ under $\gamma=0.5$ in Table \ref{tab:alphabar} to see how large $\bar{\alpha}$ could be. The numbers in Table \ref{tab:alphabar} reveal that the value of $\bar{\alpha}$ varies systemically as the ratio of $M_0$ to $M$ varies. $\bar{\alpha}$ approaches to $\gamma$ as the ratio increases to 1 and approaches to $0$ as the ratio diminishes to zero. We present the values of $\bar{\alpha}$ under $\gamma=0.25$ and $\gamma=0.75$ in \hyperref[sec:appendixA]{Appendix A}. The result implies that one cannot use conventional significance levels $\{0.01, 0.05, 0.1\}$ when the ratio $M_0/M$ exceeds 0.5.  

\begin{table}[h]
      \caption{Values of $\bar{\alpha}$'s in Theorem 1 With $M=10, 20, \ldots, 100$, $M_0=kM-1$ for $k=0.1, 0.2, \ldots, 1$, and $\gamma=0.5$. }
\begin{center}
    \begin{tabular}{cccccccccccc}
    \hline \hline
          &       & \multicolumn{10}{c}{$k$} \\
\cmidrule{3-12}    $M$   &       & 1     & 0.9   & 0.8   & 0.7   & 0.6   & 0.5   & 0.4   & 0.3   & 0.2   & 0.1 \\
    \hline
    10    &       & 0.453 & 0.378 & 0.301 & 0.227 & 0.159 & 0.101 & 0.056 & 0.024 & 0.006 & . \\
    20    &       & 0.477 & 0.403 & 0.327 & 0.253 & 0.184 & 0.123 & 0.073 & 0.036 & 0.013 & 0.001 \\
    30    &       & 0.485 & 0.412 & 0.336 & 0.262 & 0.192 & 0.130 & 0.079 & 0.041 & 0.015 & 0.002 \\
    40    &       & 0.489 & 0.416 & 0.341 & 0.266 & 0.196 & 0.134 & 0.082 & 0.043 & 0.017 & 0.003 \\
    50    &       & 0.491 & 0.418 & 0.343 & 0.269 & 0.199 & 0.136 & 0.084 & 0.045 & 0.018 & 0.003 \\
    60    &       & 0.492 & 0.420 & 0.345 & 0.270 & 0.200 & 0.138 & 0.086 & 0.046 & 0.019 & 0.004 \\
    70    &       & 0.494 & 0.421 & 0.346 & 0.272 & 0.201 & 0.139 & 0.087 & 0.046 & 0.019 & 0.004 \\
    80    &       & 0.494 & 0.422 & 0.347 & 0.273 & 0.202 & 0.140 & 0.087 & 0.047 & 0.019 & 0.004 \\
    90    &       & 0.495 & 0.423 & 0.348 & 0.273 & 0.203 & 0.140 & 0.088 & 0.047 & 0.020 & 0.004 \\
    100   &       & 0.495 & 0.423 & 0.348 & 0.274 & 0.204 & 0.141 & 0.088 & 0.048 & 0.020 & 0.004 \\
    \hline
    \end{tabular}%
\end{center}
  \label{tab:alphabar}
\end{table}

Among the three conditions postulated in Theorem \ref{prop:mains}, the first condition in Theorem \ref{prop:mains} is crucial because it prevents the distribution of the test statistics from degenerating. The condition is satisfied if the set of alternative forecasts $\mathbf{M}$ contains at least one forecast that attains the same risk as the benchmark. This condition is violated if all the alternative forecasts in $\mathbf{M}$ are strictly riskier than the benchmark forecast. In this case, both test statistics diverge to $-\infty$ while the critical values converge to some fixed numbers so the rejection probability converges to zero. Therefore, if the first condition is violated, the conclusion no longer holds.

The second condition states that the set $\mathbf{M}$ contains at least one forecast that is riskier than the benchmark. Recall that in the example from the previous subsection $\mu_{2}<0$ plays the key role drawing the conclusion by having the limiting distribution of $\hat{T}^s_n$ deviate from that of $\hat{T}^{s*}_{n,b}$. In the same manner, the second condition in the theorem causes the asymptotic distribution of the test statistics to differ from that of the bootstrap test statistics.
If the second condition is not satisfied, then $\mu$ is zero under the null hypothesis so the bootstrap test statistics correctly approximate the limiting distribution of the test statistics. The probability to reject the null hypothesis, therefore, converges to the significance level $\alpha$, rather than exceeding $\alpha$.

Unlike the first two, the last condition is not a necessary condition. It requires that the covariance of $(Z_i, Z_j)$ is zero for any $i\neq j \in \mathbf{M}$ where $Z $ is the random vector from $N(0_M, \Sigma)$ in Assumption \ref{as:weakconv}. In a simple case where $M=2$, we can easily show that the result still holds even if $\text{cov}(Z_1, Z_2) > 0.$ The condition is posited to simplify the proof.

\subsection*{3.3. Remedying pointwise asymptotic invalidity with moment selection}\label{subsec:modifiedhybrid}
As our foremost finding of this article, Theorem \ref{prop:mains} shows that the hybrid test is not pointwise asymptotically of level $\alpha$ for some $\alpha$. Lemma \ref{lemma:LFC} identifies the root of this problem. Namely, the distribution from which the limit of the two-sided critical value $\hat{c}^{s*}_n$ is obtained does not stochastically dominate the asymptotic distribution of the two-sided test statistic $\hat{T}^s_n$, i.e., $\lim_{n\to\infty}J^s_n(\cdot, \mu)$ where $J^s_n$ is defined in \eqref{eq:def_distfn}. This problem can be solved if we approximate the asymptotic distribution of $\hat{T}^s_n$ and obtain a critical value from it. The moment selection technique provides a simple way to do so. 

The moment selection technique was originally proposed by \cite{hansen2005test}. The purpose was to improve the power of the Reality Check which exploits the least favorable case, because the Reality Check tends to perform conservatively by picking a rather large critical value in the sense of \eqref{eq:lfc}. 
\cite{andrews2010inference}, \cite{canay2010}, and \cite{bugni2010} independently developed similar techniques in the context of testing moment inequality. See \cite{canayshaikh} for more details.

Our interest does not lie in improving the power of the hybrid test. Nonetheless, the moment selection technique can serve to rectify the problem that we state in Theorem \ref{prop:mains}. Below we explain how we can apply the generalized moment selection method by \cite{andrews2010inference} to the hybrid test.

First, we normalize test statistics so that their values are zero under the null hypothesis. Specifically, define two modified test statistics $\tilde{T}^r_n$ and $\tilde{T}^s_n$ as 
\begin{align}
\tilde{T}^r_n & \equiv \sqrt{n} \max_{ m \in \mathbf{M}} \left( \frac{\hat{d}_{n,m}}{\hat{\sigma}_{n,m}} \vee 0 \right)  = S^{r} \left(\hat{D}^{-\frac{1}{2}}_n \sqrt{n} \hat{d}_n \right ) \text{ and }\nonumber\\
\tilde{T}^s_n & \equiv \sqrt{n} \min \left( \max_{ m \in \mathbf{M}} \left( \frac{\hat{d}_{n,m}}{\hat{\sigma}_{n,m}} \vee 0 \right),  \max_{ m \in \mathbf{M}} \left(  \left( -\frac{\hat{d}_{n,m}}{\hat{\sigma}_{n,m}}\right) \vee 0 \right) \right) = S^{s} \left(\hat{D}^{-\frac{1}{2}}_n \sqrt{n} \hat{d}_n \right) \label{eq:modi_stat}
\end{align}
where $S^q: \mathbb{R}^M \to \mathbb{R}$ for $q \in \{r,s\}$ are real-valued functions such that $S^r (x) = \max_{m \in\mathbf{M}} (x \vee 0)$ and $S^s (x) = \min(\max_{m \in\mathbf{M}} (x \vee 0), \max_{m \in\mathbf{M}} ((-x) \vee 0))$. The operation $a \vee b$ denotes the maximum between $a$ and $b$.

Second, we define the moment selecting vector $\hat{\psi}_n = (\hat{\psi}_ {n,1}, \ldots, \hat{\psi}_ {n,M} )^t$ where its $m$th element is 
\begin{align}
\hat{\psi}_ {n,m} \equiv  \frac{\sqrt{n}}{\kappa_n} \frac{\hat{d}_{n,m}}{\hat{\sigma}_{n,m}}1\left\{\frac{\sqrt{n}}{\kappa_n} \frac{\hat{d}_{n,m}}{\hat{\sigma}_{n,m}} < - 1 \right\} \text{ for }m=1, \ldots, M. \label{eq:momentselectingvec}
\end{align}
$\kappa_n$ is a non-stochastic sequence of non-negative numbers such that $\kappa_n \to  \infty$ and $\kappa_n/ \sqrt{n} \rightarrow 0 $ as $n \to \infty$. $\kappa_n$ is a tuning parameter that a researcher has to choose. \cite{andrews2010inference} recommend $\kappa_n = \sqrt{\log n}.$

We suggest two types of data-dependent critical values. The first type is bootstrap-based. We define the critical values as follows:
\begin{align}
\tilde{c}^{q*}_n (1-\alpha) &\equiv \inf \left\{x \in \mathbf{R}: P^*_n\left\{ \tilde{T}^{q*}_{n,b}\leq x \right\} \geq 1-\alpha  \right\} \text{ for }q \in \{r, s\} \label{eq:type2cv}
\end{align}
where 
\begin{align}
 &\tilde{T}^{q*}_{n,b} \equiv  S^{q} \left(\hat{D}^{-\frac{1}{2}}_n \sqrt{n} \tilde{d}^{*}_{n,b} + \hat{\psi}_n \right) \text{ for }q \in \{r,s\} \text{ for }b=1, \ldots, B,  \label{eq:modified}
\end{align}
$P^*_n$ is the bootstrap probability conditional on the sample $\{d_t\}^n_{t=1}$ and $B$ is some large number. The second type is simulation-based. We define critical values $\tilde{c}^{q}_n (1-\alpha)$ for $q \in \{r,s\}$ as 
\begin{align}
\tilde{c}^{q}_n (1-\alpha) &\equiv \inf \left\{x \in \mathbf{R}: 
P^{\#} \left\{ 
S^q \left( \hat{\Omega}^{\frac{1}{2}}_n Z^{\#} +   \hat{\psi}_n\right) \leq x \right\} \geq 1-\alpha  
\right\}\label{eq:type1cv}
\end{align}
where $\hat{\Omega}_n \equiv \hat{D}^{-\frac{1}{2}}_n \hat{\Sigma}_n \hat{D}^{-\frac{1}{2}}_n$, $\hat{\Sigma}_n$ is an estimator of $\Sigma$ in Assumption \ref{as:weakconv}, and $\hat{\Omega}_n^{1/2}$ is a symmetric positive semi-definite matrix such that $\hat{\Omega}^{1/2}_n \hat{\Omega}^{1/2}_n=\hat{\Omega}_n$. $P^{\#}$ a probability measure conditioned on $( \hat{\Omega}^{\frac{1}{2}}_n, \hat{\psi}_n)$; $Z^{\#}$ follows the standard normal distribution and is independent from the sample $\{d_t\}^n_{t=1}$. Given $( \hat{\Omega}^{\frac{1}{2}}_n, \hat{\psi}_n)$, we can obtain $\tilde{c}^q_n(1-\alpha)$ by simulating $\{Z^{\#}_1, \ldots, Z^{\#}_R\}$ for some large $R$. 

The biggest difference of the modified bootstrap test statistics in \eqref{eq:modified} from the original bootstrap test statistics in \eqref{eq:bootstat} is that the moment selecting vector $\hat{\psi}_n$ defined in \eqref{eq:momentselectingvec} is added to the centred bootstrap statistics $\tilde{d}^*_{n,b}$. The bootstrap statistic $\tilde{d}^*_{n,b} $ hinders the bootstrap test statistics from approximating asymptotic distribution of the test statistics when $\mu$ is not zero. To be specific, if $\mu_m <0$ for some $m \in \mathbf{M}$, $\sqrt{n} \hat{d}_{n,m} /\hat{\sigma}_{n,m}$ diverges to $-\infty$ whereas $\sqrt{n}\tilde{d}^*_{n,b,m} /\hat{\sigma}_{n,m}$ remains stochastically bounded. The moment selecting vector aids the approximation by adding a quantity diverging to $-\infty$ to $\sqrt{n}\tilde{d}^*_{n,b,m}/\hat{\sigma}_{n,m}$ when $\sqrt{n}\hat{d}_{n,m}/\hat{\sigma}_{n,m}$ is sufficiently small. By the same logic, the moment selecting vector allows $S^q ( \hat{\Omega}^{\frac{1}{2}}_n Z^{\#} +   \hat{\psi}_n)$ in \eqref{eq:type1cv} to approximate the distribution of the test statistic $\tilde{T}^q_n$ for $q\in\{r,s\}$.

Given the test statistics and the critical values, we are ready to define the modified hybrid test.
\begin{definition}\label{def:modifiedhybridtest}
With the test statistics $ \tilde{T}^r_n $ and $ \tilde{T}^s_n $ defined in \eqref{eq:modi_stat}, the modified hybrid test is defined by
\begin{align*}
\tilde{\phi}_n  \equiv 1\{ \tilde{T}^r_n > c^r_n(1-\alpha(1-\gamma)) \text{ or }\tilde{T}^s_n > c^s_n(1-\alpha\gamma) \} 
\end{align*}
for any $\gamma \in (0,1)$ where $(c^r_n, c^s_n) = (\tilde{c}^{r*}_n,\tilde{c}^{s*}_n)$ in \eqref{eq:type2cv} or $(c^r_n, c^s_n) = (\tilde{c}^r_n,\tilde{c}^s_n)$ in \eqref{eq:type1cv}.
\end{definition}

Because both types of critical values use an estimator for $\Sigma$, we strengthen Assumption \ref{as:consist_var} and require consistency of $\hat{\Sigma}$. 
\begin{assumption}\label{as:consistent_Sigma}
$\hat{\Sigma}_n \overset{p}{\to}\Sigma$ as $n\to\infty.$
\end{assumption}
\noindent With this reinforced assumption, we define $\mathcal{F}^{pt}$ as
\begin{align*}
    \mathcal{F}^{pt}\equiv  \{(\Sigma, F):\text{Assumption \ref{as:stationarity}, \ref{as:weakconv}, \ref{as:bconsistency}, \ref{as:psd_cov} and \ref{as:consistent_Sigma} are satisfied}\}.
\end{align*}
Since Assumption \ref{as:consistent_Sigma} implies Assumption \ref{as:consist_var}, we have $\mathcal{F}^{pt} \subset \mathcal{F}.$ We define the subset of $\mathcal{F}^{pt}$ that satisfies the null hypothesis as $\mathcal{F}^{pt}_0$, i.e.,  $\mathcal{F}^{pt}_0 \equiv \{(\Sigma, F)\in \mathcal{F}^{pt}: E[d_{t}]\leq 0\}$, over which we show pointwise asymptotic validity of the modified hybrid test.

\begin{proposition} \label{prop:modifiedtest}
Suppose Assumption \ref{as:stationarity}, \ref{as:weakconv}, \ref{as:bconsistency}, \ref{as:psd_cov} and \ref{as:consistent_Sigma} hold. Let $\kappa_n$ be a sequence such that 
\begin{align*}
    \kappa_n \to \infty \text{ and } \frac{\kappa}{\sqrt{n}} \to 0\text{ as }n\to \infty.
\end{align*}
Let $\alpha \in (0,0.5)$ and $\gamma \in (0,1)$. Then, for $(c^r_n, c^s_n) = (\tilde{c}^r_n,\tilde{c}^s_n)$ or $(c^r_n, c^s_n) = (\tilde{c}^{r*}_n,\tilde{c}^{s*}_n)$, the modified hybrid test in Definition \ref{def:modifiedhybridtest} satisfies
\begin{align*}
\limsup_{n \to \infty} E[\tilde{\phi_n}] \leq \alpha  \text{ for any }(\Sigma, F) \in \mathcal{F}^{pt}_0.
\end{align*} 
\end{proposition}
Proposition \ref{prop:modifiedtest} shows that the modified hybrid test is pointwise asymptotically of level $\alpha$ within the class of data generating processes, $\mathcal{F}^{pt}$. That is, that given a data generating process $(\Sigma, F)\in \mathcal{F}^{pt}_0$, one could carry out the modified hybrid test while keeping the probability of committing the Type I error less than the significance level $\alpha$ for large $n$. 

Though the modified hybrid test adopts the general moment selection approach, the result in \cite{andrews2010inference} does not directly apply to our setup. They show the proposed test enjoys uniform asymptotic validity, which is a necessary condition of pointwise asymptotic validity and we formally define in \hyperref[sec:appendixC]{Appendix C}. Monotonicity of their test statistic plays a crucial role in attaining uniform asymptotic validity as pointed out by \cite{canayshaikh}. However, the two-sided test statistic $\tilde{T}^s_n$ of the modified hybrid test is not monotone in $\hat{d}_n$ and thus violates Assumption 1(a) and 3 in \cite{andrews2010inference}. Therefore, we alter their proof and achieve pointwise asymptotic validity. The proof of Proposition \ref{prop:modifiedtest} can be found in \hyperref[sec:appendixB]{Appendix B}.

Uniform asymptotic validity is a stronger condition than pointwise asymptotic validity in the sense that the former implies the latter. Due to the complication derived from the fact that SPA tests intrinsically deal with dependent data such as time series, the literature of SPA tests has developed focusing on pointwise asymptotic validity in contrast to the moment inequality tests in which the importance of uniform asymptotic validity has been addressed. The main focus of this paper is neither to propose an alternative to the hybrid test and nor to investigate theoretical properties of the modified hybrid test. Nonetheless, we provide a small example showing that the modified hybrid test does not satisfy uniform asymptotic validity under conditions used by \cite{andrews2010inference} in \hyperref[sec:appendixC]{Appendix C}. 

\section*{{4. Monte Carlo simulation}}\label{sec:montecarlo}
While Theorem \ref{prop:mains} tells us that the hybrid test could be pointwise asymptotically invalid, it doesn't inform us how pronounced the distortion could be in a finite sample. In this section, we explore how significantly pointwise asymptotic invalidity manifests in a finite sample through Monte Carlo simulation. Furthermore, we study the finite sample rejection probabilities of the modified hybrid test under the null hypothesis.

We use the simulation design similar to those considered in \cite{song2012testing} and \cite{hansen2005test}. As in \hyperref[sec:setup]{Section 2}, suppose we have a benchmark forecast and $M$ distinct alternative forecasts. We observe $n$ realized relative risks between the benchmark forecast $\varphi_{t,0}$ and $m$th alternative forecast $\varphi_{t,m}$, $d_{t,m}$ for $t=1, \ldots, n$ and $m=1,\ldots,M$. We are interested in testing the null hypothesis in \eqref{eq:hypotheses} meaning that the benchmark is superior to all the alternative forecasts in terms of expected risk.

For simulation, we draw realized relative risks independently from a normal distribution, i.e., $d_{t} \sim \text{i.i.d.}~ N(- \lambda_{M_0}, V)$ where $\lambda_{M_0}$ is an $M$ dimensional vector of which the first $M_0$ elements are zeros and the rest $M-M_0$ elements are ones. $M_0$ refers to the number of the alternative forecasts of which expected risks are the same as that of the benchmark as before. The relative risk $\mu =-c \lambda_k$ is non-positive and hence the design satisfies the null hypothesis. The i.i.d. observations imply that Assumption \ref{as:stationarity}, \ref{as:weakconv} and \ref{as:bconsistency} are satisfied. The variance-covariance matrix $V$ is designed to satisfy the third condition of Theorem \ref{prop:mains}. The off-diagonal elements of the variance matrix $V$ are zeros and the $M$ diagonal elements are determined by a random draw from a uniform distribution over $[1,2]$ at the beginning of the simulation and are fixed during the simulation.


The sample size $n$ is 200 and we draw $M \times n$ random numbers. The number of Monte Carlo repetitions and the bootstrap samples are 5,000 and 500 respectively. For the number of alternative forecasts, we consider $M \in \{50, 100\}$. For the significance level, we consider 0.01, 0.05, and 0.10. We use $\gamma=0.5$ as recommended by \cite{song2012testing}, and $\kappa_n = \sqrt{\log n}$ for the tuning parameter in the modified hybrid test as recommended by \cite{andrews2010inference}.

Table \hyperref[tab:MCsimu]{2} reports the simulated rejection probabilities. Hyb. indicates the hybrid test in \eqref{eq:hybridtest} while Boot. and Simu. refer to the modified hybrid test in Definition \hyperref[def:modifiedhybridtest]{1} with the bootstrap-based and simulation-based critical values in \eqref{eq:type2cv} and \eqref{eq:type1cv} respectively. 

Table \hyperref[tab:MCsimu]{2} provides evidence supporting Theorem \ref{prop:mains}. Many simulated rejection probabilities of the hybrid test exceed the significance level $\alpha$ when $M_0$ is strictly less than $M$. This phenomenon is the most pronounced when $M_0$ is slightly less than $M$, and the extent of distortion is not marginal. For example, the rejection probabilities of the hybrid test with $M=50$ and $M_0=45$ are 0.208, 0.149, and 0.070 which are almost twice, three times, and seven times larger than their corresponding significance levels $\alpha=0.10, 0.05$ and $0.01$. We have similar results in the case with $M=100$ and $M_0=95.$

Furthermore, there is a noticeable pattern in the simulated probabilities. First, when all inequalities are binding, that is, when the second condition in Theorem \ref{prop:mains} is not satisfied, the probabilities are close to the nominal level $\alpha$. This is because, under this data generating process, the bootstrap distribution correctly approximates the distribution of the test statistics and hence the rejection probability converges exactly to the nominal level. Second, as $M_0$ decreases, the probabilities abruptly increase, exceeding the nominal level $\alpha$ but decline gradually. This is because both test statistics converge to $\max_{m \in M_0} Z_m$, which decreases in $M_0$, where $\{Z_m: m=1, \ldots, M\}$ are independent random variables from the standard normal distribution. On the contrary, the limiting distributions of the bootstrap test statistics do not depend on $M_0$. This difference leads to diminishing rejection probabilities along decreasing $M_0$. Finally, the probabilities fall below $\alpha$ when the ratio $M_0/M$ is small: less than 0.4 for $\alpha=0.10$, 0.3 for $\alpha=0.05$, and 0.2 for $\alpha=0.01$ for the case $M=50.$ This is consistent with our findings from Table \ref{tab:alphabar} that $\bar{\alpha}$ decreases as the ratio $M_0/M$ diminishes.

Contrary to the hybrid test and as expected from our result in Proposition \ref{prop:modifiedtest}, the simulated rejection probabilities of the modified hybrid tests are less than the nominal level except only two cases with $M=M_0=50$ and $\alpha=0.01$. The modified hybrid test appears to be conservative in that the simulated rejection probabilities are close to $\alpha/2$ when $M_0$ is strictly less than $M$. This is because two test statistics $\tilde{T}^r_n$ and $ \tilde{T}^s_n$ converge in distribution to the same distribution to which $\hat{T}^r_n$ and $ \hat{T}^s_n$ converge. Furthermore, comparing the simulated probabilities from `Boot.' and `Simu' tells us that two different critical values of the modified hybrid test, one bootstrap-based and the other simulation-based yield similar results.

\begin{table}[htbp]\label{tab:MCsimu}%
\begin{threeparttable}
 \caption{Simulated Rejection Probabilities}
           \begin{tabular}{cccccccccccccc}
    \hline\hline
          &       &       & \multicolumn{3}{c}{$\alpha=0.10$} &       & \multicolumn{3}{c}{$\alpha=0.05$} &       & \multicolumn{3}{c}{$\alpha=0.01$} \\
\cmidrule{4-6}\cmidrule{8-10}\cmidrule{12-14}    $M$   & $M_0$ &       & Hyb.  & Boot. & Simu. &       & Hyb.  & Boot. & Simu. &       & Hyb.  & Boot. & Simu. \\
    \hline
    50    & 50    &       & 0.106 & 0.083 & 0.083 &       & 0.055 & 0.046 & 0.043 &       & 0.016 & 0.011 & 0.011 \\
    50    & 45    &       & 0.208 & 0.056 & 0.057 &       & 0.149 & 0.029 & 0.028 &       & 0.070 & 0.006 & 0.007 \\
    50    & 40    &       & 0.192 & 0.052 & 0.053 &       & 0.139 & 0.029 & 0.027 &       & 0.060 & 0.005 & 0.005 \\
    50    & 35    &       & 0.164 & 0.052 & 0.051 &       & 0.113 & 0.024 & 0.026 &       & 0.050 & 0.005 & 0.005 \\
    50    & 30    &       & 0.139 & 0.053 & 0.053 &       & 0.095 & 0.028 & 0.028 &       & 0.047 & 0.007 & 0.007 \\
    50    & 25    &       & 0.115 & 0.052 & 0.052 &       & 0.086 & 0.030 & 0.028 &       & 0.038 & 0.007 & 0.006 \\
    50    & 20    &       & 0.102 & 0.062 & 0.060 &       & 0.074 & 0.033 & 0.033 &       & 0.036 & 0.008 & 0.008 \\
    50    & 15    &       & 0.075 & 0.052 & 0.052 &       & 0.051 & 0.028 & 0.026 &       & 0.024 & 0.007 & 0.006 \\
    50    & 10    &       & 0.053 & 0.055 & 0.055 &       & 0.036 & 0.027 & 0.026 &       & 0.016 & 0.008 & 0.007 \\
    50    & 5     &       & 0.025 & 0.054 & 0.054 &       & 0.016 & 0.026 & 0.025 &       & 0.006 & 0.005 & 0.004 \\
          &       &       &       &       &       &       &       &       &       &       &       &       &  \\
    100   & 100   &       & 0.103 & 0.081 & 0.083 &       & 0.053 & 0.042 & 0.041 &       & 0.011 & 0.009 & 0.008 \\
    100   & 95    &       & 0.219 & 0.062 & 0.061 &       & 0.157 & 0.033 & 0.032 &       & 0.073 & 0.010 & 0.008 \\
    100   & 90    &       & 0.225 & 0.062 & 0.061 &       & 0.162 & 0.030 & 0.031 &       & 0.073 & 0.007 & 0.006 \\
    100   & 85    &       & 0.208 & 0.065 & 0.064 &       & 0.147 & 0.035 & 0.032 &       & 0.068 & 0.008 & 0.008 \\
    100   & 80    &       & 0.190 & 0.060 & 0.060 &       & 0.137 & 0.032 & 0.032 &       & 0.059 & 0.009 & 0.009 \\
    100   & 75    &       & 0.185 & 0.060 & 0.061 &       & 0.133 & 0.030 & 0.029 &       & 0.057 & 0.007 & 0.008 \\
    100   & 70    &       & 0.172 & 0.057 & 0.058 &       & 0.116 & 0.029 & 0.029 &       & 0.052 & 0.008 & 0.008 \\
    100   & 65    &       & 0.155 & 0.0572 & 0.057 &       & 0.111 & 0.028 & 0.027 &       & 0.048 & 0.007 & 0.007 \\
    100   & 60    &       & 0.147 & 0.057 & 0.058 &       & 0.102 & 0.030 & 0.030 &       & 0.044 & 0.007 & 0.006 \\
    100   & 55    &       & 0.136 & 0.053 & 0.053 &       & 0.090 & 0.029 & 0.029 &       & 0.038 & 0.008 & 0.007 \\
    \hline
    \end{tabular}%
 \begin{tablenotes}
 \item NOTE: Hyb., Boot., and Simu. refer to the hybrid test and modified hybrid test with bootstrap-based and simulation-based critical values respectively. 
 \end{tablenotes} 
\end{threeparttable}
\end{table}%

\section*{{5. Conclusion}}\label{sec:conclusion}
\cite{song2012testing} proposes the hybrid test but does not formally discuss its theoretical properties. In this article, we demonstrate with a simple example that the hybrid test may be not pointwise asymptotically of level $\alpha$ at commonly used significance levels, and provide a formal result generalizing this observation. Pointwise asymptotic invalidity of the hybrid test has a practical implication in that a researcher may commit the Type I error with probability larger than $\alpha$ even with a large sample. As an easy fix, we propose a modified hybrid test by adjusting the generalized moment selection approach to our setup, which is often used in the literature of testing moment inequality. We prove that the modified hybrid test enjoys pointiwse asymptotic validity. Finally we present Monte Carlo results supporting the theoretical findings.  

While SPA tests and moment inequality tests share many common features, the literature of the former has pursued pointwise asymptotic validity and that of the latter has addressed the importance of uniform asymptotic validity. We attribute the reason to the complex nature of dependent data generating processes that SPA tests deal with. Modifying the hybrid test so as to gain uniform asymptotic validity or developing a general framework for uniform asymptotic validity in the context of SPA tests could be an interesting topic, yet it is beyond the scope of this paper and we leave it for future research.





\bibliography{song}
\bibliographystyle{agsm}

\newpage

This supplementary file consists of three appendices. \hyperref[sec:appendixA]{Appendix A} presents the tables for the values of $\bar{\alpha}$'s in Theorem \ref{prop:mains} under $\gamma=0.25$ and $\gamma=0.75$. \hyperref[sec:appendixB]{Appendix B} provides auxiliary lemmas and proofs for the main results in \hyperref[sec:body]{Section 3}. Finally, we define uniform asymptotic validity and show that the modified hybrid test is not uniformly asymptotically of level $\alpha$ in \hyperref[sec:appendixC]{Appendix C}.

\section*{Appendix A: Values of $\bar{\alpha}$ under various $\gamma$ }
\label{sec:appendixA}
\setcounter{table}{0}
\renewcommand{\thetable}{A.\arabic{table}} 

\begin{table}[!htbp]
      \caption{The values of $\bar{\alpha}$'s in Theorem 1 with $M=10, 20,30, \ldots, 100$, $M_0=kM-1$ for $k=0.1, 0.2, \ldots, 1$, and $\gamma=0.25$. }
\begin{center}
        
    \begin{tabular}{cccccccccccc}
    \hline\hline
          &       & \multicolumn{10}{c}{$k$} \\
\cmidrule{3-12}    $M$   &       & 1     & 0.9   & 0.8   & 0.7   & 0.6   & 0.5   & 0.4   & 0.3   & 0.2   & 0.1 \\
    \hline
    10    &       & 0.225 & 0.182 & 0.142 & 0.106 & 0.074 & 0.047 & 0.026 & 0.012 & 0.003 & . \\
    20    &       & 0.238 & 0.195 & 0.155 & 0.118 & 0.085 & 0.057 & 0.034 & 0.017 & 0.006 & 0.001 \\
    30    &       & 0.242 & 0.199 & 0.159 & 0.122 & 0.089 & 0.060 & 0.037 & 0.019 & 0.007 & 0.001 \\
    40    &       & 0.244 & 0.201 & 0.161 & 0.124 & 0.090 & 0.062 & 0.038 & 0.020 & 0.008 & 0.001 \\
    50    &       & 0.245 & 0.203 & 0.162 & 0.125 & 0.092 & 0.063 & 0.039 & 0.021 & 0.009 & 0.002 \\
    60    &       & 0.246 & 0.203 & 0.163 & 0.126 & 0.092 & 0.063 & 0.040 & 0.022 & 0.009 & 0.002 \\
    70    &       & 0.246 & 0.204 & 0.164 & 0.126 & 0.093 & 0.064 & 0.040 & 0.022 & 0.009 & 0.002 \\
    80    &       & 0.247 & 0.204 & 0.164 & 0.127 & 0.093 & 0.064 & 0.040 & 0.022 & 0.009 & 0.002 \\
    90    &       & 0.247 & 0.205 & 0.164 & 0.127 & 0.094 & 0.065 & 0.041 & 0.022 & 0.009 & 0.002 \\
    100   &       & 0.248 & 0.205 & 0.165 & 0.127 & 0.094 & 0.065 & 0.041 & 0.022 & 0.009 & 0.002 \\
    \hline
    \end{tabular}%

    \end{center}
\end{table}

\begin{table}[h]
      \caption{The values of $\bar{\alpha}$'s in Theorem 1 with $M=10, 20,30, \ldots, 100$, $M_0=kM-1$ for $k=0.1, 0.2, \ldots, 1$, and $\gamma=0.75$. }
\begin{center}
    \begin{tabular}{cccccccccccc}
    \hline\hline
          &       & \multicolumn{10}{c}{$k$} \\
\cmidrule{3-12}    $M$   &       & 1     & 0.9   & 0.8   & 0.7   & 0.6   & 0.5   & 0.4   & 0.3   & 0.2   & 0.1 \\
    \hline
    10    &       & 0.691 & 0.598 & 0.490 & 0.374 & 0.263 & 0.166 & 0.090 & 0.038 & 0.009 & . \\
    20    &       & 0.722 & 0.636 & 0.533 & 0.420 & 0.306 & 0.203 & 0.119 & 0.058 & 0.020 & 0.002 \\
    30    &       & 0.732 & 0.648 & 0.547 & 0.435 & 0.321 & 0.216 & 0.130 & 0.066 & 0.024 & 0.004 \\
    40    &       & 0.736 & 0.654 & 0.554 & 0.442 & 0.328 & 0.223 & 0.135 & 0.070 & 0.027 & 0.005 \\
    50    &       & 0.739 & 0.658 & 0.558 & 0.447 & 0.332 & 0.227 & 0.138 & 0.072 & 0.028 & 0.005 \\
    60    &       & 0.741 & 0.660 & 0.561 & 0.450 & 0.335 & 0.229 & 0.141 & 0.074 & 0.029 & 0.006 \\
    70    &       & 0.742 & 0.662 & 0.563 & 0.452 & 0.337 & 0.231 & 0.142 & 0.075 & 0.030 & 0.006 \\
    80    &       & 0.743 & 0.663 & 0.564 & 0.453 & 0.339 & 0.233 & 0.143 & 0.076 & 0.030 & 0.006 \\
    90    &       & 0.744 & 0.664 & 0.566 & 0.455 & 0.340 & 0.234 & 0.144 & 0.076 & 0.031 & 0.006 \\
    100   &       & 0.745 & 0.665 & 0.566 & 0.456 & 0.341 & 0.235 & 0.145 & 0.077 & 0.031 & 0.007 \\
    \hline
    \end{tabular}%
    \end{center}
\end{table}

\newpage
\section*{Appendix B: Auxiliary lemma and proofs}\label{sec:appendixB}
Before we proceed, it is useful to introduce some additional notation that we use throughout this section. Let $Z$ be a $M$-dimensional normal vector from $N(0_M, \Sigma)$ where $\Sigma$ is the $M\times M$ covariance matrix in Assumption \ref{as:weakconv}. That is,
\begin{align}
    \sqrt{n}\hat{D}^{-1/2}_n(\hat{d}_n-\mu) \overset{d}{\to} Z \equiv (Z_1, \ldots, Z_M)^t
    \label{eq:weakconv}
\end{align} 
as $n\to\infty$. Let $\sigma^{2}_m =\Sigma_{m,m}$, $D= \text{diag}(\Sigma)$, and $\hat{D}_n = \text{diag}(\hat{\sigma}_{n,1}, \ldots, \hat{\sigma}_{n,M})$ so that $\hat{D}_n \overset{p}{\to} D$ by Assumption \ref{as:consist_var} or \ref{as:consistent_Sigma}. $\mathbf{M}_0$ denotes the set of indices with zero mean, i.e. 
\begin{align}
\mathbf{M}_0=\{m \in \mathbf{M}: \mu_m =0 \}. \label{eq:contactset}    
\end{align}
Furthermore, let us define a function $f: \mathbb{R}^{M} \mapsto  \mathbb{R}^2$ by 
\begin{align}
    f(x) =  \left( 
    \begin{matrix}
    \max_{m \in \mathbf{M}} x_m\\
    \min(\max_{m \in \mathbf{M}} x_m, \max_{m \in \mathbf{M}} (-x_m))
    \end{matrix}
    \right) \label{eq:function}
\end{align}
where $x=(x_1, \ldots, x_{M})^t$. Clearly, both mappings $x \mapsto \max_{m \in \mathbf{M}} x_m$ and $x \mapsto \min(\max_{m \in \mathbf{M}} x_m$, $\max_{m \in \mathbf{M}} (-x_m))$ are continuous. This implies that function $f$ is also continuous in $x$. The vector of test statistics can be written as $(\hat{T}^r_n,\hat{T}^s_n)^t =f(\sqrt{n}\hat{D}^{-1/2}_n\hat{d}_n)$.

\subsection*{B.1. Auxiliary Lemma}
\begin{lemma}\label{lemma:aux1}
Suppose that Assumption \ref{as:stationarity}, \ref{as:weakconv}, \ref{as:psd_cov} and \ref{as:consist_var} hold. Let $2\leq M<\infty$. Assume that $\Sigma = I_M$.
\begin{itemize}
\item[\textit{(i)}] If $F$ satisfies $H_0$ and there is $m \in \mathbf{M}$ such that $\mu_m=0$ and $m' \in \mathbf{M}$ such that $\mu_{m'}<0$, then 
\begin{align*}
    (\hat{T}^r_n,\hat{T}^s_n) \overset{d}{\to} (\max_{m \in {\mathbf{M}_0}}Z_m,\max_{m \in {\mathbf{M}_0}}Z_m) \text{ as }n \to \infty
\end{align*}
where $\mathbf{M}_0$ is defined in \eqref{eq:contactset}.

\item[\textit{(ii)}] If $F$ satisfies $H_0$ and $\mu=0_M$, then 
\begin{align*}
    (\hat{T}^r_n,\hat{T}^s_n) \overset{d}{\to} (T^r(Z), T^s(Z)) \text{ as }n \to \infty.
\end{align*}
Furthermore, we have
\begin{align*}
    P\{T^s(Z) \leq x \} = \left\{ 
    \begin{matrix}
    2 \Phi^M(x) - (2\Phi(x) - 1)^M  & \text{ if }x \geq 0\\
     2 \Phi^M(x)  & \text{ if }x <0
    \end{matrix}
    \right.
\end{align*}
\end{itemize}
\end{lemma}

\begin{proof}
First, we consider the setting in \textit{(i)}. Then both sets $\mathbf{M}\backslash\mathbf{M}_0$ and $\mathbf{M}_0$ are not empty.  We want to show $(\hat{T}^r_n,\hat{T}^s_n)^t \overset{d}{\to} (\max_{m \in {\mathbf{M}_0}}Z_m,\max_{m \in {\mathbf{M}_0}}Z_m)^t$. The definition of weak convergence implies that we need to show
\begin{align}
   \lim_{n \to \infty} \left| E\left[ g(f(\sqrt{n}\hat{D}^{-1/2}_n\hat{d}_n))\right] - E\left[ g\left( \begin{matrix}
     \max_{m \in {\mathbf{M}_0}}Z_m\\
     \max_{m \in {\mathbf{M}_0}}Z_m
     \end{matrix}
     \right)\right]\right| =0  \label{eq:portmanteau}
\end{align}
for any bounded continuous function $g: \mathbb{R}^2 \mapsto \mathbb{R}$. 

To this end, we define two events $E_1$ and $E_2$ by
\begin{align*}
E_1 \equiv \{\min_{m \in \mathbf{M}_0}\frac{\hat{d}_{n,m}}{\hat{\sigma}_{n,m}} < \max_{m \in \mathbf{M}\backslash\mathbf{M}_0}\frac{\hat{d}_{n,m}}{\hat{\sigma}_{n,m}}\}\text{ and }
E_2 \equiv \{\max_{m \in \mathbf{M}_0}\frac{\hat{d}_{n,m}}{\hat{\sigma}_{n,m}}  > \max_{m \in \mathbf{M}\backslash\mathbf{M}_0}-\frac{\hat{d}_{n,m}}{\hat{\sigma}_{n,m}}\}.
\end{align*}
Then by rearranging the terms and subtracting $\max_{m \in \mathbf{M}\backslash\mathbf{M}_0}{\mu_m}/{{\sigma}_m}$ on both sides in the event $E_1$ we have 
\begin{align*}
    \lim_{n\to \infty}P\{E_1\}
    =\lim_{n \to \infty} 
    P\{-\max_{m \in \mathbf{M}\backslash\mathbf{M}_0}\frac{\mu_m}{{\sigma}_m}
     < 
     - \min_{m \in \mathbf{M}_0}\frac{\hat{d}_{n,m}}{\hat{\sigma}_{n,m}}
     + \max_{m \in \mathbf{M}\backslash\mathbf{M}_0}\frac{\hat{d}_{n,m}}{\hat{\sigma}_{n,m}}
     - \max_{m \in \mathbf{M}\backslash\mathbf{M}_0}\frac{\mu_m}{{\sigma}_m}\}=0.
\end{align*}
The last equality holds from the fact that 
\begin{align*}
    \min_{m \in \mathbf{M}_0}\frac{\hat{d}_{n,m}}{\hat{\sigma}_{n,m}} \overset{p}{\to} 0 \text{ and }\max_{m \in \mathbf{M}\backslash\mathbf{M}_0}\frac{\hat{d}_{n,m}}{\hat{\sigma}_{n,m}}-\max_{m \in \mathbf{M}\backslash\mathbf{M}_0}\frac{\mu_m}{{\sigma}_m} \overset{p}{\to} 0
\end{align*}
as $n\to \infty$, which are implied by Assumption \ref{as:weakconv} and continuous mapping theorem. Similarly we can show that $\lim_{n \to \infty} P\{E_2\}=0$. Let $1_{E_j}$ denote an indicator function which takes 1 as its value if the event $E_j$ occurs for $j \in \{1,2\}$ and zero otherwise. 

Then \eqref{eq:portmanteau} holds by the following. For any bounded continuous function $g: \mathbb{R}^2 \mapsto \mathbb{R}$, we have
\begin{align}
     & \lim_{n \to \infty} \left| E\left[ g(f(\sqrt{n}\hat{D}^{-1/2}_n\hat{d}_n)) -  
     g\left( \begin{matrix}
     \max_{m \in {\mathbf{M}_0}}Z_m\\
     \max_{m \in {\mathbf{M}_0}}Z_m
     \end{matrix}
     \right)
     \right]\right| \nonumber \\
&\leq  \lim_{n \to \infty} \left| E\left[ \{g(f(\sqrt{n}\hat{D}^{-1/2}_n\hat{d}_n)) - 
     g\left( \begin{matrix}
     \max_{m \in {\mathbf{M}_0}}Z_m\\
     \max_{m \in {\mathbf{M}_0}}Z_m
     \end{matrix}
     \right)\} 
     \{ 1_{E_1} + (1-1_{E_1})1_{E_2}\}
     \right]\right| \nonumber\\
     &  + \lim_{n \to \infty} \left| E\left[ \{g(f(\sqrt{n}\hat{D}^{-1/2}_n\hat{d}_n)) -  
     g\left( \begin{matrix}
     \max_{m \in {\mathbf{M}_0}}Z_m\\
     \max_{m \in {\mathbf{M}_0}}Z_m
     \end{matrix}
     \right)\}
     \{ 1-1_{E_1}\}\{ 1-1_{E_2}\}\right]\right|  \nonumber\\
     &\leq\lim_{n \to \infty} 2\sup_{x \in \mathbb{R}}|g(x)| \{P\{E_1\}+P\{E_2\}\} \nonumber \\
     & + \lim_{n \to \infty} \left| E\left[ \{g(f(\sqrt{n}\hat{D}^{-1/2}_n\hat{d}_n)) -  g\left( \begin{matrix}
     \max_{m \in {\mathbf{M}_0}}Z_m\\
     \max_{m \in {\mathbf{M}_0}}Z_m
     \end{matrix}
     \right)\}
     \{ 1-1_{E_1}\} \{ 1-1_{E_2}\}\right]\right| \nonumber\\
     &=\lim_{n \to \infty} \left| E\left[ \{g\left( \begin{matrix}
     \max_{m \in {\mathbf{M}_0}}{\sqrt{n}\hat{d}_{n,m}}/{\hat{\sigma}_{n,m}}\\
     \max_{m \in {\mathbf{M}_0}}{\sqrt{n}\hat{d}_{n,m}}/{\hat{\sigma}_{n,m}}
     \end{matrix}
     \right)
     -  g\left( \begin{matrix}
     \max_{m \in {\mathbf{M}_0}}Z_m\\
     \max_{m \in {\mathbf{M}_0}}Z_m
     \end{matrix}
     \right)\}
     \{ 1-1_{E_1}\} \{ 1-1_{E_2}\}\right]\right| \label{eq:teststat}\\
     &\leq\lim_{n \to \infty} \left| E\left[ g\left( \begin{matrix}
     \max_{m \in {\mathbf{M}_0}}{\sqrt{n}\hat{d}_{n,m}}/{\hat{\sigma}_{n,m}}\\
     \max_{m \in {\mathbf{M}_0}}{\sqrt{n}\hat{d}_{n,m}}/{\hat{\sigma}_{n,m}}
     \end{matrix}
     \right)
     - g\left( \begin{matrix}
     \max_{m \in {\mathbf{M}_0}}Z_m\\
     \max_{m \in {\mathbf{M}_0}}Z_m
     \end{matrix}
     \right)\right]\right| \nonumber\\
& + \lim_{n \to \infty}\left| E\left[
     \{g\left( \begin{matrix}
     \max_{m \in {\mathbf{M}_0}}{\sqrt{n}\hat{d}_{n,m}}/{\hat{\sigma}_{n,m}}\\
     \max_{m \in {\mathbf{M}_0}}{\sqrt{n}\hat{d}_{n,m}}/{\hat{\sigma}_{n,m}}
     \end{matrix}
     \right)
     -  g\left( \begin{matrix}
     \max_{m \in {\mathbf{M}_0}}Z_m\\
     \max_{m \in {\mathbf{M}_0}}Z_m
     \end{matrix}
     \right)\}
     \{ 1_{E_1}+1_{E_2}-1_{E_1}1_{E_2}\}\right]\right| \nonumber\\
& \leq \lim_{n \to \infty} E\left[ \left|
     g\left( \begin{matrix}
     \max_{m \in {\mathbf{M}_0}}{\sqrt{n}\hat{d}_{n,m}}/{\hat{\sigma}_{n,m}}\\
     \max_{m \in {\mathbf{M}_0}}{\sqrt{n}\hat{d}_{n,m}}/{\hat{\sigma}_{n,m}}
     \end{matrix}
     \right)
     -  
     g\left( \begin{matrix}
     \max_{m \in {\mathbf{M}_0}}Z_m\\
     \max_{m \in {\mathbf{M}_0}}Z_m
     \end{matrix}
     \right) \right| \cdot
	 \left| 1_{E_1}+1_{E_2}-1_{E_1}1_{E_2}\right|\right] \nonumber\\
     &\leq \lim_{n \to \infty} 2\sup_{x \in \mathbb{R}}|g(x)| \{P\{E_1\}+P\{E_2\}\}=0. \nonumber
\end{align}
The first inequality holds by the triangular inequality. To get the first equality we use the fact that $g$ is a bounded function, and that $(1-1_{E_1})1_{E_2} \leq  1_{E_2}$. The second equality holds from that the probabilities of two events $E_1$  and $E_2$ converge to zero, and that 
\begin{align*}
f(\sqrt{n} \hat{D}^{-1/2}_n \hat{d}_n) = \left(
     \max_{m \in {\mathbf{M}_0}}{\sqrt{n}\hat{d}_{n,m}}/{\hat{\sigma}_{n,m}},
     \max_{m \in {\mathbf{M}_0}}{\sqrt{n}\hat{d}_{n,m}}/{\hat{\sigma}_{n,m}}
   \right)^t
\end{align*}
conditional on the event $E^c_1 \cap E^c_2$. The second inequality holds by the triangular inequality again. The penultimate inequality holds by continuous mapping theorem and by the definition of weak convergence. For the last inequality, we bound $g$ with its supremum. The last equality again holds by that the probability  of two events converge to zero. Therefore, the vector of test statistics $(\hat{T}^r_n, \hat{T}^s_n)$ has a desired limit distribution.

Next, we consider \textit{(ii)}. Because $\mu_m=0$ for at least an element of $\mu$, $\mathbf{M}_0$ is not empty. The weak convergence result holds as a result of continuous mapping theorem. The formula for the distribution function can be derived from the following:
  \begin{align}
            &P\{T^s(Z) > x \} \nonumber\\
            &=P( \min(\max_{m \in \mathbf{M}} Z_m, -\min_{m \in \mathbf{M}} Z_m)> x) \nonumber\\
            &= P(\max_{m \in \mathbf{M}} Z_m> x, \text{ and } -\min_{m \in \mathbf{M}} Z_m> x) \nonumber\\
            &= 1-P(\max_{m \in \mathbf{M}} Z_m \leq  x, \text{ or } \min_{m \in \mathbf{M}} Z_m \geq - x) \nonumber\\
            &=1-\{P(\max_{m \in \mathbf{M}} Z_m \leq  x) + P(\min_{m \in \mathbf{M}} Z_m \geq - x) - P(\forall m \in \mathbf{M}, - x \leq  Z_m \leq  x)\} \label{eq:RHS}\\
            &= \left\{ 
            \begin{matrix}
            1-2\Phi(x)^M + (2\Phi(x)-1)^M 	& \text{ if }x \geq 0 \\
            1-2\Phi(x)^M\quad 				&\text{ if }x<0.
				\end{matrix}\right.             \nonumber
        \end{align}
To get the last equality, we use that $\Phi(x) = 1-\Phi(-x)$ and that $P(-x < Z_m < x)=0$ for any $m$ if $x<0$.
\end{proof}

\subsection*{B.2. Proof for Lemma \ref{lemma:LFC}}
Let $F$ in Assumption \ref{as:stationarity} be the distribution which satisfies the null hypothesis and Assumptions \ref{as:weakconv} and \ref{as:psd_cov}. First, the only constraint imposed for $\Sigma$ is that it should be a positive definite $M\times M$ matrix. We consider $F$ with $\Sigma=I_M.$ Second, we consider $\mu=(\mu_{1}, \ldots, \mu_{M})^t$ such that $\mu_{1}<0$ and $\mu_m=0$ for $m=2,\ldots,M.$

Under such $F$, the asymptotic distribution of $T^s(\hat{D}^{-1/2}_n \sqrt{n}(\hat{d}_n-\mu))$ is identical to $T^s(Z)$ by Lemma \ref{lemma:aux1}\textit{(ii)}. That is,
\begin{align*}
    \lim_{n \to \infty} J^s_n(x;0_M) = \left\{ 
            \begin{matrix}
            2\Phi(x)^M - (2\Phi(x)-1)^M 	& \text{ if }x \geq 0 \\
            2\Phi(x)^M\quad 				&\text{ if }x<0
				\end{matrix}\right.     
\end{align*}
where $J^s_n$ is defined in \eqref{eq:def_distfn}. Next, the asymptotic distribution of $T^s_n$ under $F$ is $\max_{m\in \mathbf{M}_0}Z_m$ with $\mathbf{M}_0=\{2,\ldots, M\}$ as given in Lemma \ref{lemma:aux1}\textit{(i)} and so
\begin{align*}
    \lim_{n \to \infty} J^s_n(x;\mu) = P\{\max_{m\in \mathbf{M}_0}Z_m \leq x\} = \Phi(x)^{M-1}.
\end{align*}

If $T^s(\hat{D}^{-1/2}_n \sqrt{n}(\hat{d}_n-\mu))$ stochastically dominates $T^s_n$ in the limit, then it must be that $\lim_{n \to \infty} J^s_n(x;0_M)  \leq \lim_{n \to \infty} J^s_n(x;\mu)$ for all $x\in \mathbb{R}$. However, for $x>0$, it holds that
\begin{align*}
     &\lim_{n \to \infty} J^s_n(x;\mu) - \lim_{n \to \infty} J^s_n(x;0_M)= (2\Phi(x)-1) \{(2\Phi(x)-1)^{M-1} - \Phi(x)^{M-1}\} < 0
\end{align*}
where the last strict inequality holds because $2\Phi(x)-1>0$ and $2\Phi(x)-1 < \Phi(x)$ for $x>0$. Hence we conclude.

\subsection*{B.3. Proof for Theorem \ref{prop:mains}}
This proof consists of three parts. In the first two parts we obtain the probability limits of critical values $\hat{c}^{r*}_n$ and $\hat{c}^{s*}_n$ respectively. Then in the last part we show the existence of $\bar{\alpha}$ which satisfies the conclusion.

$(\Sigma, F) \in \mathcal{F}_0$ is the data generating process and hence $\mu=E[d_t] \leq 0_M$. Note that by the first two conditions, both sets $\mathbf{M}\backslash\mathbf{M}_0$ and $\mathbf{M}_0$ are not empty where $\mathbf{M}_0$ is defined in \eqref{eq:contactset}. 

\paragraph{Part 1:} We start from obtaining the probability limits of the critical value  $\hat{c}^{s*}_n$. Assumption \ref{as:bconsistency} states that 
\begin{align*}
    \sup_{x \in \mathbb{R}^{M}} \left| P^*_n\{\sqrt{n}(\hat{d}^*_{n,b} - \hat{d}_n)\leq x\} - P\{\sqrt{n} (\hat{d}_n-\mu) \leq x\} \right| \overset{p}{\to} 0
\end{align*}
as $n$ diverges to infinity where $P^*_n$ denotes the bootstrap probability measure. Define a random vector $(Z^r, Z^s)^t \equiv f(Z)$ where $f$ is defined in \eqref{eq:function}. Then the continuous mapping theorem implies that 
\begin{align}
    \sup_{x, y \in \mathbb{R}} \left| 
    P^*_n\{\hat{T}^{r*}_{n,b} \leq x, \hat{T}^{s*}_{n,b} \leq y \} -
    P\{Z^r \leq x, Z^s \leq y\} \right| \overset{p}{\to} 0
    \label{eq:bconsistency}
\end{align}
as $n\to \infty$. Since the mapping that selects a coordinate $(x, y) \mapsto y$ is continuous, we have 
\begin{align*}
\sup_{y \in \mathbb{R}} \left|     P^*_n\{\hat{T}^{s*}_{n,b}  \leq y \} -P\{Z^s \leq y\} \right| \overset{p}{\to} 0 \text{ as }n\to\infty
\end{align*}
by Theorem 10.8 of \cite{kosorok2008}. Lemma \ref{lemma:aux1} provides the distribution function of $Z^s = T^s(Z)$ and it is continuous and strictly increasing. Given this, Lemma 11.2.1 of \cite{lehmann2006testing} gives
\begin{align}
    \hat{c}^{s*}_n \overset{p}{\to} c^s \equiv \inf\{y \in \mathbb{R}: P\{Z^s\leq y\} \geq 1-\alpha\gamma\} \text{ as }n\to\infty \label{eq:limitcv1}
\end{align}
for any $\alpha \in (0,1) $ and $\gamma \in (0,1]$. 

\paragraph{Part 2:} For the probability limit of the one-sided critical value $\hat{c}^{r*}_n$, we only need to consider the case $\gamma \in (0,1)$ because $\hat{c}^{r*}_n=\infty$ if $\gamma=1.$ We start with showing that 
\begin{align}
    \sup_{x \in \mathbb{R}} \left| G_n(x) \right| 
    \equiv\sup_{x \in \mathbb{R}} \left| P^*_n\{
    \hat{T}^{r*}_{n,b}  1\{\hat{T}^{s*}_{n,b} \leq \hat{c}^{s*}_n\} \leq x \} - P\{Z^r 1\{Z^s\leq c^s\}\leq x\}\right| \overset{p}{\to} 0. 
    \label{eq:goal0}
\end{align}
 The term on the left-hand-side can be bounded as follows:
\begin{align}\label{eq:goal1}
&\sup_{x \in \mathbb{R}} \left|G_n(x)\right|\nonumber \\
&\leq \sup_{x <0 } \left|G_n(x)\right|
+ \sup_{x \geq 0 } \left| G_n(x)\right| \nonumber \\
&= \sup_{x <0 } \left| P^*_n\{\hat{T}^{r*}_{n,b} \leq x, \hat{T}^{s*}_{n,b} \leq \hat{c}^{s*}_n \} - P\{Z^r \leq x, Z^s \leq c^s\}\right| 
+ \sup_{x \geq 0 } \left| G_n(x)\right| \nonumber\\
&= \sup_{x <0 } \left| P^*_n\{\hat{T}^{r*}_{n,b}  \leq x, \hat{T}^{s*}_{n,b} \leq \hat{c}^{s*}_n \} - P\{Z^r \leq x, Z^s \leq c^s\}\right| \nonumber \\
& +\sup_{x \geq 0 } \left| P^*_n\{\hat{T}^{r*}_{n,b}  \leq x, \hat{T}^{s*}_{n,b} \leq \hat{c}^{s*}_n \} +P^*_n\{\hat{T}^{s*}_{n,b} > \hat{c}^{s*}_n\} - P\{Z^r \leq x, Z^s \leq c^s\} - P\{ Z^s > c^s\}\right| \nonumber\\
 &\leq 2\sup_{x \in \mathbb{R}} \left|P^*_n\{\hat{T}^{r*}_{n,b}  \leq x, \hat{T}^{s*}_{n,b} \leq \hat{c}^{s*}_n\} - P\{Z^r \leq x, Z^s \leq c^s\}\right|  \\
 &\quad+ \left|P^*_n\{ \hat{T}^{s*}_{n,b} > \hat{c}^{s*}_n\} - P\{ Z^s > c^s\} \right|  \nonumber \\
 &\leq 2\sup_{x \in \mathbb{R}} \left|
        P^*_n\{\hat{T}^{r*}_{n,b}  \leq x, \hat{T}^{s*}_{n,b} \leq \hat{c}^{s*}_n \} - 
        P\{Z^r  \leq x, Z^s \leq  \hat{c}^{s*}_n\}
        \right| \nonumber\\
    &\quad + 2\sup_{x \in \mathbb{R}} \left|
        P\{Z^r  \leq x, Z^s \leq \hat{c}^{s*}_n \} - 
        P\{Z^r  \leq x, Z^s \leq  c^s \}
        \right| \nonumber\\
    &\quad    + \left|P^*_n\{ \hat{T}^{s*}_{n,b} > \hat{c}^{s*}_n \} - P\{ Z^s > c^s\} \right| . \nonumber 
\end{align}
The first inequality holds by the triangular inequality. The first equality follows from the fact that for $\hat{T}^{r*}_{n,b} 1\{\hat{T}^{s*}_{n,b} \leq \hat{c}^{s*}_n\}$ to take a negative value the indicator function must be one. Similarly, we get the second equality by decomposing $G_n(x)$ into two cases where the indicator function is zero or not. For the second inequality we use that the supremum is a non-decreasing set operator and the last inequality holds by the triangular inequality. 

Now we show that all three terms in the last line of \eqref{eq:goal1} are asymptotically negligible. The convergence of the first term comes from that 
\begin{align*}
&\sup_{x \in \mathbb{R}} \left|
        P^*_n\{\hat{T}^{r*}_{n,b}  \leq x, \hat{T}^{s*}_{n,b} \leq \hat{c}^{s*}_n \} - 
        P\{Z^r  \leq x, Z^s \leq  \hat{c}^{s*}_n \}
        \right|
\\ 
&\leq 
\sup_{x, y \in \mathbb{R}} \left|
        P^*_n\{\hat{T}^{r*}_{n,b}  \leq x, \hat{T}^{s*}_{n,b} \leq y \} - 
        P\{Z^r  \leq x, Z^s \leq  y\}
        \right|
\end{align*}
and (\ref{eq:bconsistency}). To show the convergence of the second term, define the joint distribution function of $(Z^r , Z^s)$ and the marginal distribution function of $Z^s$ by 
\begin{align*}
F_{rs} (x,y) \equiv P\{Z^r  \leq x, Z^s \leq y \} \text{ and }F_s (y) \equiv P\{Z^s \leq y \}.
\end{align*}
Both functions are continuous. The continuous mapping theorem and consistency of critical value $\hat{c}^{s*}_n$ imply the pointwise convergence,
\begin{align*}
        F_{rs} (x,\hat{c}^{s*}_n)-F_{rs} (x,c^s)
          \overset{p}{\to} 0\text{ for every }x \in \mathbb{R}.
\end{align*}
The same logic gives the pointwise convergence of conditional distribution function,
\begin{align*}
\frac{F_{rs} (x,\hat{c}^{s*}_n)}{ F_s (\hat{c}^{s*}_n)}  - \frac{F_{rs} (x,c^s)}{ F_s (c^s)} \overset{p}{\to} 0 \text{ for every }x \in \mathbb{R}.
\end{align*}
Now we can extend this pointwise convergence into the uniform convergence over the real-line by applying Theorem 11.2.9 of \cite{lehmann2006testing} to two conditional distributions as the conditional distribution function ${F_{rs} (x,c^s)}/{ F_s(c^s)}$ is continuous.  Once we have the uniform convergence of the conditional distributions, we have
\begin{align*}
&\sup_{x \in \mathbb{R}} \left|
        P\{Z^r  \leq x, Z^s \leq \hat{c}^{s*}_n \} - 
        P\{Z^r  \leq x, Z^s \leq  c^s \}
        \right|\\ 
&= \sup_{x \in \mathbb{R}} \left|
        F_{rs}(x, \hat{c}^{s*}_n ) - F_{rs}(x, c^s) 
        \right| \\
&\leq\sup_{x \in \mathbb{R}}  \left|
        \frac{F_{rs}(x, \hat{c}^{s*}_n)}{F_s(\hat{c}^{s*}_n)} - \frac{F_{rs} (x,c^s)}{ F_s (c^s)} 
        \right|\cdot F_s(\hat{c}^{s*}_n) + \sup_{x \in \mathbb{R}}  \left|\frac{F_{rs} (x,c^s)}{ F_s (c^s)} \right|\cdot \left|F_s (\hat{c}^{s*}_n)-F_s (c^s) \right|
        \overset{p}{\to} 0.
\end{align*}
The convergence of the third term is straightforward.

Given this result, let us obtain the probability limit of critical value $\hat{c}^{r*}_n$. We can't directly apply Lemma 11.2.1 of \cite{lehmann2006testing} as in Step 1 because the distribution of $Z^r  1\{Z^s \leq c^s\} $ is discontinuous at zero. Let $\alpha \in (0,1- 2^{-M})$. Then we have
\begin{align}\label{eq:restriction on alpha}
    P\{Z^r  1\{Z^s \leq c^s\} \leq 0\}
    &= P\{Z^r  \leq 0, Z^s \leq c^s\} + P\{Z^s > c^s\} \nonumber\\
    &= P\{Z^r  \leq 0, Z^s \leq c^s\} + \alpha\gamma \nonumber\\
    &\leq \min(P\{Z^r  \leq 0\},P\{ Z^s \leq c^s\}) + \alpha\gamma \\
    &=\min(2^{-M}, 1-\alpha\gamma) + \alpha\gamma \nonumber\\
    &< 1-\alpha  + \alpha \gamma  = 1-\alpha(1-\gamma) \quad \text{ if }\alpha <1-2^{-M}.\nonumber
\end{align}
The second equality holds by the definition of $c^s$. The third equality holds by that $P\{Z^r  \leq 0 \}=P\{Z_m \leq 0 \text{ for all }m\in \mathbf{M}\} = \Phi^{M}(0)=2^{-M}$ and again by the definition of $c^s$ where $\Phi(\cdot)$ is the cumulative distribution function of the standard normal distribution. This result guarantees that the $1-\alpha(1-\gamma)$ quantile of $Z^r  1\{Z^s \leq c^s\}$ is strictly positive given that $\alpha <1-2^{-M}$. As the distribution function $P\{Z^r  1\{Z^s \leq c^s\} \leq x\}$ is continuous and strictly increasing over the interval $[0, \infty)$, we have that 
\begin{align}
     \hat{c}^{r*}_n \overset{p}{\to} c^r \equiv \inf\{x \in \mathbb{R}: P\{Z^r 1\{Z^s \leq c^s\} \leq x\} \geq 1-\alpha(1-\gamma)\} \label{eq:limitcv2}
\end{align}
by Lemma 11.2.1 of \cite{lehmann2006testing}.

\paragraph{Part 3:} We show that there exists $\bar{\alpha}$ which makes the probability to reject the null hypothesis strictly greater than $\alpha$ for all $\alpha \in (0, \bar{\alpha})$. 

First, we compute a lower bound for the limiting rejection probability. The test function $\phi_n$ is defined by 
\begin{align*}
    \phi_n \equiv 1\{\hat{T}^{s}_n > \hat{c}^{s*}_n\} + 1\{\hat{T}^{s}_n \leq \hat{c}^{s*}_n\}1\{\hat{T}^{r}_n > \hat{c}^{r*}_n\}.
\end{align*}
Given the distribution $F$, for $\gamma \in (0,1)$ the limiting rejection probability is 
        \begin{align}
            \lim_{n \to \infty} E[\phi_n]
            = \lim_{n \to \infty} P\{\hat{T}^{s}_n> \hat{c}^{s*}_n \text{ or }\hat{T}^{r}_n > \hat{c}^{r*}_n\}
            = P\{\max_{m \in \mathbf{M}_0}Z_m > \min (c^s, c^r)\}. \label{eq:LHS}
        \end{align}
This holds by the weak convergence result in (\ref{eq:portmanteau}), by convergence of the critical values in (\ref{eq:limitcv1}) and (\ref{eq:limitcv2}), and by the Slutsky theorem. Define 
\begin{align}
k\equiv k(\alpha)\equiv \Phi(c^s)=1-\Phi(-c^s).  \label{eq:step4k}
\end{align}
Note that $k$ is a function of $\alpha$ as $c^s$ depends on $\alpha$. The limiting rejection probability in (\ref{eq:LHS}) is bounded from below by $1-k$ because it holds that 
        \begin{align*}
            P\{ \max_{m \in \mathbf{M}_0}Z_m > \min (c^s, c^r)\}
            \geq P\{ \max_{m \in \mathbf{M}_0}Z_{m} > c^s \}
            = 1- k^{M_0}
        \end{align*}
where $M_0 = |\mathbf{M}_0| \geq 1$. For $\gamma=1$, $\hat{c}^{r*}_n = \infty$ so we have the same lower bound for $ \lim_{n \to \infty} E[\phi_n]$ because 
\begin{align*}
     \lim_{n \to \infty} E[\phi_n]
        = \lim_{n \to \infty} P\{\hat{T}^{s}_n> \hat{c}^{s*}_n \}
        =P\{ \max_{m \in \mathbf{M}_0}Z_{m} > c^s \}
        = 1- k^{M_0}.
\end{align*}
Therefore in order to attain the conclusion, it is sufficient to find $\alpha$ satisfying that $1- k^{M_0}> \alpha$. 

Now let us consider the relationship between $k$ defined in \eqref{eq:step4k} and $\alpha$. The definition of $c^s$ provides the connection between the two. Lemma \ref{lemma:aux1}\textit{(ii)} and the definition of $c^s$ give
        \begin{align}
            \alpha\gamma 
            &= P(Z^s > c^s) \nonumber\\
            &= \left\{ 
            \begin{matrix}
            1-2k^M + (2k-1)^M 	& \text{ if }c^s \geq 0 \text{ or equivalently if }\alpha\gamma \in (0, 0.5]\\
            1-2k^M\quad 				&\text{ if }c^s<0 \text{ or equivalently if }\alpha\gamma \in (0.5, 1).
				\end{matrix}\right.             \label{eq:RHS}
        \end{align}
        
We consider the case $\alpha \in (0, 0.5/\gamma]$. Recall that the tuning parameter $\gamma \in (0, 1]$ is fixed. Following (\ref{eq:restriction on alpha}), define a function $a_{\gamma}: [0, 1] \to [0,\frac{1}{\gamma}]$ by
\begin{align*}
a_{\gamma}(x)\equiv \left\{
\begin{matrix}
\frac{1}{\gamma}(1-2x^M +(2x-1)^M) & \text{ if } x \in [0.5, 1]\\
\frac{1}{\gamma} (1-2x^M)& \text{ if } x \in [0, 0.5)
\end{matrix}
\right. .
\end{align*}
It is easy to check that $a_{\gamma}$ is continuous on $[0, 1]$ and $a'_{\gamma}(s)<0$ for all $x \in (0,1)$. This implies that $a_{\gamma}$ is bijective. In other words, for $k \in [0, 1]$ there exists one-to-one relation between $k$ and $\alpha$, and $a_{\gamma}$ is the inverse function of $k(\alpha)$.

Given the finding, let's obtain the set of values for $\alpha$ satisfying $1-k^{M_0}>\alpha$. Specifically, find the values of $x\in [0.5, 1]$ satisfying the following condition:
        \begin{align*}
            h_{\gamma}(x) \equiv 1-x^{M_0} - a_{\gamma}(x)  >0
        \end{align*}
where $h_{\gamma}$ is a real-valued function defined on $[0, 1]$. It is easy to check $h_{\gamma}(1)=0$ and $\lim_{x\to 1-}h'_{\gamma}(x)<0$. Since $h_{\gamma}$ is a polynomial, there exists $\bar{\varepsilon} \in (0, 0.5)$ satisfying that $h_{\gamma}(x)>0$ for all $x \in (1-\bar{\varepsilon} , 1)$. Note that it's not trivial to obtain a closed-form solution for $\bar{\varepsilon}$ because it is the solution to the $M$th degree polynomial equation. However, for fixed $M$ and $M_0$, the value of $\bar{\varepsilon} $ can be numerically approximated and so is $\bar{\alpha}.$

Therefore any value $\alpha$ in the interval $(0, a_{\gamma}(1-\bar{\varepsilon}))$ satisfies $1- k^{M_0} > \alpha$. Recall that (\ref{eq:restriction on alpha}) in Step 3 requires $\alpha$ to be less than $1-2^{-M}$. As a result, we have the desired result by setting $\bar{\alpha} = \min(a_{\gamma}(1-\bar{\varepsilon}), 1-2^{-M}, (2\gamma)^{-1})$.

\subsection*{B.4 Proof of Proposition \ref{prop:modifiedtest}}
In this section, we prove Proposition \ref{prop:modifiedtest} by modifying the proof of Lemma 2 in \cite{andrews2010inference}. As addressed in the main text, their results do not apply directly to our setup. The two-sided test statistic $\tilde{T}^s_n$ violates Assumption 1(a) and Assumption 3 in \cite{andrews2010inference}. Their Assumption 1(a) requires that the test statistics must be monotone in $\hat{d}_n$, but our statistic $\tilde{T}^s_n$ is not monotone in $\hat{d}_n$ as the function $S^s(\cdot)$ in \eqref{eq:modi_stat} is weakly increasing in $x$ if $x \leq 0$ and weakly decreasing if $x\geq 0$. Their Assumption 3 requires $S^s(x)$ to be strictly positive if and only if $x_m>0$ for some $m \in \mathbf{M}$. It's easy to show that $S^s(x)=0$ if $M=2$, $x_1>0$ and $x_2>0.$ 

Our aim in this proposition is, however, to prove pointwise asymptotic validity which is a weaker condition than uniform asymptotic validity that \cite{andrews2010inference} show. The fact that we deal with a fixed data generating process $(\Sigma, F)$ allows us to apply the generalized moment selection technique even without Assumption 1(a) and Assumption 3 in their paper. As in \cite{andrews2010inference}, we prove the case for using the simulation-based critical values $\tilde{c}^q_n(1- \alpha)$ for $q \in \{r,s\}$ defined in \eqref{eq:type1cv}. The other case with the bootstrap critical values $\tilde{c}^{q*}_n(1- \alpha)$ for $q \in \{r,s\}$ defined in \eqref{eq:type2cv} can be shown in a similar manner. 

Before we begin, we define some additional notation. Define a vector ${\mu}^*=({\mu}^*_{1}, \ldots,{\mu}^*_{M})^t$ such that 
\begin{align*}
{\mu}^*_{m} = \left\{ 
\begin{matrix}
- \infty & \text{ if }\mu_{m}  <0\\
0& \text{ if }\mu_{m}  =0
\end{matrix}
\right. \text{ for }m=1, \ldots, M.
\end{align*}
Define a function $\psi: \mathbb{R}^M \to [-\infty, 0]$ such that $\psi(\xi) =(\psi_1(\xi), \ldots, \psi_M(\xi))^t$ and 
\begin{align*}
\psi_m (\xi) = \left\{ 
\begin{matrix}
\xi_m & \text{ if }\xi_{m}  < -1\\
0		   & \text{ if }\xi_{m}  \geq -1
\end{matrix}
\right. \text{ for }m=1, \ldots, M
\end{align*}
where $\xi_m $ is the $m$th element of $\xi \in \mathbb{R}^M$. Then the moment selecting vector $\hat{\psi}_n$ in \eqref{eq:momentselectingvec} can be written as $\psi( \hat{\xi}_n)$ where $\hat{\xi}_n \equiv \kappa^{-1}_n \sqrt{n} \hat{D}^{-1/2}_n \hat{d}_n$. Given the definition of $\psi(\cdot)$ and $\mu^*$, define distribution function 
\begin{align}
L^q (x) \equiv P \{ S^q (\Omega^{1/2}_0 Z^{\#} + \psi(\mu^*)) \leq x\}    \label{eq:dist_Lq}
\end{align}
for $x \in \mathbb{R}$ and $q\in\{r,s\}$ where $\Omega_0$ is the probability limit of $\hat{\Omega}_n = \hat{D}^{-1/2}_n \hat{\Sigma}_n\hat{D}^{-1/2}_n$ which is the $M\times M$ asymptotic correlation matrix of $\sqrt{n}(\hat{d}_n - \mu)$ and where $Z^{\#}$ and $S^q$ are defined as in \eqref{eq:type1cv}. Let $c^q_{\mu^*}(1-\alpha)$ be the $1-\alpha$ quantile from $L^q$. For convenience we drop the dependence on $ 1-\alpha$ and use $\tilde{c}^q_n$ and $c^q_{\mu^*}$ instead of $\tilde{c}^q_n(1-\alpha)$ and $c^q_{\mu^*}(1-\alpha)$. Furthermore, superscript $q$ implies $q$ could be substituted with $r$ or $s$. For example, we proceed with $L^q$ without explaining that $q \in \{r,s\}$ unless necessary for brevity.

The following proof consists of three steps. In the first step, we show that the critical value $\tilde{c}^q_n(1-\alpha)$ defined in \eqref{eq:type1cv} converges in probability to $c^q_{\mu^*}(1-\alpha)$ as $n\to \infty$. In the next step, we derive the weak convergence of the test statistics for the modified hybrid test, i.e., $\tilde{T}^q_n  \overset{d}{\to} S^q(\Omega_0^{1/2} Z^{\#} + \psi(\mu^*))$. In the last step, using the Slutsky theorem we combine the two results from the previous steps and derive the conclusion on pointwise asymptotic validity. 

\paragraph{Step 1: }Since the bootstrap test statistics are non-negative, the critical value $\tilde{c}^q_n$ is always non-negative as well. In order to show $\tilde{c}^q_n \overset{p}{\to} c^q_{\mu^*}$ as $n\to \infty$, we focus on the case where $c^q_{\mu^*}>0$ and later consider the other case where $c^q_{\mu^*}=0$ in the final step. Note that the condition that $c^q_{\mu^*}>0$ implies that $\mu^* \neq (-\infty)^M $ because $c^q_{\mu^*}$ should be zero by its definition if $\mu^*= (-\infty)^M$. 

We start with showing
\begin{align}
\hat{\xi}_n \equiv \kappa^{-1}_n \sqrt{n} \hat{D}^{-1/2}_n \hat{d}_n \overset{p}{\to}    \mu^* \text{ as }n\to \infty. \label{eq:conv_hat_xi}
\end{align}
If $\mu_m=0$, then by Assumption \ref{as:weakconv}, \ref{as:psd_cov} and \ref{as:consistent_Sigma} as $n\to\infty$ 
\begin{align*}
    \hat{\xi}_{n,m} = \frac{\sqrt{n}}{\kappa_n} \frac{1}{\hat{\sigma}_{n,m}} \hat{d}_{n,m} =\frac{1}{\kappa_n} \frac{\sigma_m}{\hat{\sigma}_{n,m}} \sqrt{n} \frac{\hat{d}_{n,m} - \mu_m}{\sigma_m} = \frac{1}{\kappa_n}(1+ o_p(1)) O_p(1)\overset{p}{\to} 0=\mu^*_m. 
\end{align*}
Similarly if $\mu^*_m=- \infty$ then 
\begin{align*}
    \hat{\xi}_{n,m} 
    &=\frac{1}{\kappa_n} \frac{\sigma_m}{\hat{\sigma}_{n,m}} \sqrt{n} \frac{\hat{d}_{n,m} - \mu_m}{\sigma_m} + \frac{\sqrt{n}}{\kappa_n} \frac{\sigma_m}{\hat{\sigma}_{n,m}} \frac{\mu_m}{\sigma_m}= o_p(1)- \frac{\sqrt{n}}{\kappa_n}|O_p(1)| \to -\infty =\mu^*_m
\end{align*}
with probability approaching 1. 

Next, we argue that for a sequence $\xi$
\begin{align}
 \psi(\xi) \to \psi(\mu^*) \text{ if }   \xi \to \mu^*. \label{eq:conv_xi}
\end{align}
If $\mu^*_m=0$ for some $m\in \mathbf{M}$, then $\psi_m(\xi) \to \psi_m(\mu^*)$ because $\psi_m$ is continuous at zero. If $\mu^*_m = -\infty$ for some $m\in \mathbf{M}$, then $\psi_m(\mu^*) = -\infty$ because $\xi$ eventually becomes smaller than $-1$ and for such $\xi$ it holds that $\psi_m(\xi) = \xi_m$. 

For any deterministic sequence such that $(\xi, \Omega) \to (\mu^*, \Omega_0)$, \eqref{eq:conv_xi} and continuity of $S^q$ give
\begin{align*}
S^q (\Omega^{1/2} Z^{\#} + \psi(\xi)) \to S^q (\Omega^{1/2}_0 Z^{\#} + \psi(\mu^*)) \text{ almost surely in }[Z^{\#}]
\end{align*}
where $Z^{\#} \sim N(0, I_M)$. From this, we derive that for any $x>0$
\begin{align*}
1\{ S^q (\Omega^{1/2} Z^{\#} + \psi(\xi)) \leq x \} \to 1\{ S^q (\Omega^{1/2}_0 Z^{\#} + \psi(\mu^*)) \leq x \} \text{ almost surely in }[Z^{\#}].
\end{align*}
This holds because for any $x>0$
\begin{align*}
    P\{ S^q (\Omega^{1/2} Z^{\#} + \psi(\mu^*)) =x\}=0 
\end{align*}
because $\mu^* \neq (-\infty)^M$ and the distribution of $ S^q (\Omega^{1/2}_0 Z^{\#} + \psi(\mu^*))$ is strictly increasing and continuous for $x>0$. Then the dominated convergence theorem provides
\begin{align}
P\{ S^q (\Omega^{1/2} Z^{\#} + \psi(\xi)) \leq x \} \to P\{ S^q (\Omega^{1/2}_0 Z^{\#} + \psi(\mu^*)) \leq x \} \label{eq:contisq}
\end{align}
for any $x >0$. This implies that $ P \{ S^q (\Omega^{1/2} Z^{\#} + \psi(\xi)) \leq x\}$ is continuous function in $(\xi, \Omega)$ at $(\mu^*, \Omega_0).$ In sum, \eqref{eq:contisq}, \eqref{eq:conv_hat_xi} and the continuous mapping theorem combine to give
\begin{align*}
\left| P^{\#} \{ S^q (\hat{\Omega}^{1/2}_n Z^{\#} + \psi(\hat{\xi}_n)) \leq x\} - P \{ S^q (\Omega^{1/2}_0 Z^{\#} + \psi(\mu^*)) \leq x\}\right| \overset{p}{\to} 0
\end{align*}
for any $x>0$ as $n\to \infty$ where $P^{\#}$ denotes the conditional probability given $(\hat{\xi}_n, \hat{\Omega}_n)$. Recall that $\tilde{c}^q_n$ in \eqref{eq:type1cv} is the $1-\alpha$ quantile of $P^{\#} \{ S^q (\hat{\Omega}^{1/2}_n Z^{\#} + \psi(\hat{\xi}_n)) \leq x\}$ and $c^q_{\mu^*}$ is the $1-\alpha$ quantile of $L^q$ defined in \eqref{eq:dist_Lq}. Because we consider the case where $ c^q_{\mu^*}>0 $, we have $\tilde{c}^q_n \overset{p}{\to} c^q_{\mu^*}$ by Lemma 5 of \cite{AB2010}. 

\paragraph{Step 2: }In this step, we derive the weak convergence of the test statistics, $$\tilde{T}^q_n  \overset{d}{\to} S^q(\Omega_0^{1/2} Z^{\#} + \psi(\mu^*)) \text{ as }n\to \infty.$$ To do so, we borrow arguments in \cite{andrews2009validity} that they use to verify Assumption B0. If any element in $\mu$ is strictly 
negative, then $\hat{D}^{-1/2}_n \sqrt{n}\hat{d}_n$ does not converge in distribution and the continuous mapping theorem cannot be applied. To circumvent this problem we consider a function of $\hat{D}^{-1/2}_n \sqrt{n}\hat{d}_n$ which converges in distribution regardless of any value of $\mu \leq 0.$ Let $G(\cdot)$ be a strictly increasing continuous distribution function on $\mathbb{R}$ such as the distribution function of the standard normal distribution $\Phi$. Define
\begin{align*}
    g_{n,m} \equiv G\left(\frac{\sqrt{n}\hat{d}_{n,m}}{\hat{\sigma}_{n,m}}\right) = G\left(\frac{\sigma_{m}}{\hat{\sigma}_{n,n}} \left( \frac{\sqrt{n} (\hat{d}_{n,m} - \mu_m)}{\sigma_m}\right) + \frac{\sigma_{m}}{\hat{\sigma}_{n,n}}  \frac{\sqrt{n}\mu_m}{\sigma_m}\right)
\end{align*}
for each $m \in \mathbf{M}$. Assumption \ref{as:weakconv}, \ref{as:psd_cov}, \ref{as:consistent_Sigma} and the continuous mapping theorem imply
\begin{align*}
    g_{n,m} \overset{d}{\to} g(\frac{Z_m}{\sigma_m} + \mu^*_m) \text{ for }m \in \mathbf{M}
\end{align*}
 where $Z\sim N(0_M, \Sigma)$ in \eqref{eq:weakconv} and $G(-\infty)=0$. These results hold jointly and combine to give 
 \begin{align}
     g_n \equiv (g_{n,1}, \ldots, g_{n,M}) \overset{d}{\to} (g(\frac{Z_1}{\sigma_1} + \mu^*_1), \ldots, g(\frac{Z_M}{\sigma_M} + \mu^*_M) ) \equiv g_{\infty}. \label{eq:weakconv_g} 
 \end{align}
Let $G^{-1}$ denote the inverse of $G.$ For $x=(x_1, \ldots, x_M)^t \in (\mathbb{R}\cup \{-\infty\})^M$ let 
$$\tilde{G}(x)=(G(x_1), \ldots, G(x_M)) \in [0,1)^M.$$
For $y=(y_1, \ldots, y_M)^t \in [0,1)^M$ let $$\tilde{G}^{-1}(y) = (G^{-1} (y_1), \ldots,G^{-1} (y_M)) \in (\mathbb{R}\cup \{-\infty\})^M.$$ Define $S^{q*}$ as
\begin{align*}
    S^{q*} (y) \equiv  S^{q}(\tilde{G}^{-1}(y)) \text{ for }y\in [0,1)^M.
\end{align*}
Then the weak convergence of the test statistic follows from the following:
\begin{align*}
    \tilde{T}^q _n 
    &= S^{q}(\tilde{G}^{-1}(g_n)) 
    = S^{q*} (g_n) \overset{d}{\to} S^{q*} (g_{\infty})
    = S^{q}(\tilde{G}^{-1} (g_{\infty})) 
    = S^{q}(D^{-1/2}Z + \mu^*).
\end{align*}
The first and the third equality hold by the definition of $\tilde{G}^{-1}$ and $g_n$, the second equality holds by the definition of $S^{q*}$, and the convergence holds by \eqref{eq:weakconv_g} and the continuous mapping theorem.

Finally note that $S^q(\Omega^{1/2}_0 Z^{\#} + \psi(\mu^*))$ has the same distribution as $ S^{q}(D^{-1/2}Z + \mu^*)$ because $\psi(\mu^*) = \mu^*$ and $\Omega^{1/2}_0 Z^{\#}  \sim N(0_M, \Omega_0)$ and $D^{-1/2}Z \sim N(0_M, D^{-1/2}\Sigma D^{-1/2})$ where $\Omega_0= D^{-1/2}\Sigma D^{-1/2}$.

\paragraph{Step 3: }We finally derive the result on pointwise asymptotic validity. In Step 1 and 2, we considered the case where $c^q_{\mu^*}>0$. For such case, the results from the two step and the Slutsky theorem yield \begin{align*}
\liminf_{n \to \infty} P \{ \tilde{T}^q_{n} \leq \tilde{c}^q_{n}(1-\alpha) \} 
= P\{ S^q (\Omega^{1/2}_0 Z^{\#} + \psi(\mu^*)) \leq c^q_{\mu^*}(1-\alpha)\} \geq 1-\alpha
\end{align*}
where the second inequality holds because $c^q_{\mu^*}(1-\alpha)$ is the $1-\alpha$ quantile of $L^q$ in \eqref{eq:dist_Lq}. As this result holds both for $q \in \{r,s\}$, by sub-additivity of the probability measure we have
\begin{align}&
\limsup_{n \to \infty} P\{ \tilde{T}^r_{n} > \tilde{c}^r_{n}(1-\alpha(1-\gamma)) \text{ or } \tilde{T}^s_{n} > \tilde{c}^s_{n}(1-\alpha\gamma) \}\nonumber \\
&\leq \limsup_{n \to \infty}  \left\{ P \{\tilde{T}^r_{n} > \tilde{c}^r_{n}(1-\alpha(1-\gamma)) \} + P \{\tilde{T}^s_{n} > \tilde{c}^s_{n}(1-\alpha\gamma) \} \right\} \leq  \alpha. \label{eq:propfinal}
\end{align}

Finally we derive the conclusion still holds even if $c^q_{\mu^*}(1-\alpha)=0$ for all $\alpha\in (0,1/2)$. To this end, note that 
\begin{align*}
P \{\tilde{T}^r_{n} \leq \tilde{c}^r_{n}(1-\alpha(1-\gamma))\}
&\geq P \{\tilde{T}^r_{n} \leq c^r_{\mu^*}(1-\alpha(1-\gamma))\} \\ 
&=P \{ \hat{D}^{-1/2}_{n} \sqrt{n} \hat{d}_{n} \leq  c^r_{\mu^*}(1-\alpha(1-\gamma))\}\\
&\to P\{D^{-1/2} Z + \mu^* \leq c^r_{\mu^*}(1-\alpha(1-\gamma))\}\\
&= P\{ \Omega^{1/2}_0 Z^{\#} + \mu^* \leq  c^r_{\mu^*}(1-\alpha(1-\gamma)) \}\\
&=P\{ S^r(\Omega^{1/2}_0 Z^{\#} + \mu^*) \leq  c^r_{\mu^*}(1-\alpha(1-\gamma)) \} \geq 1-\alpha(1-\gamma)
\end{align*}
where the first inequality holds because $\tilde{c}^r_{n}(1-\alpha)$ is non-negative; the first equality holds by the definition of $\tilde{T}^r_n$; the convergence holds by Assumption \ref{as:weakconv}, \ref{as:psd_cov} and \ref{as:consistent_Sigma}; the next equality holds because $D^{-1/2}Z + \mu^* \overset{d}{=} \Omega^{1/2}_0 Z^{\#} + \mu^*$; the last equality holds by the definition of $S^r$; and the last inequality holds by definition of $c^r_{\mu^*}(1-\alpha(1-\gamma))$. Similarly, we have
\begin{align*}
P \{\tilde{T}^s_{n} \leq \tilde{c}^s_{n}(1-\alpha\gamma)\}
&\geq P \{\tilde{T}^s_{n} \leq c^s_{\mu^*}(1-\alpha\gamma)\} \\ 
&=P \{ \hat{D}^{-1/2}_{n} \sqrt{n} \hat{d}_{n} \leq  c^s_{\mu^*}(1-\alpha\gamma) \text{ or } - \hat{D}^{-1/2}_{n} \sqrt{n} \hat{d}_{n} \leq  c^s_{\mu^*}(1-\alpha) \}\\
&\to P\{ \Omega^{1/2}_0 Z^{\#} + \mu^* \leq  c^s_{\mu^*}(1-\alpha\gamma)  \text{ or } -\Omega^{1/2}_0 Z^{\#} + \mu^* \leq  c^s_{\mu^*}(1-\alpha\gamma) \}\\
&=P\{ S^s(\Omega^{1/2}_0 Z^{\#} + \mu^*) \leq  c^s_{\mu^*}(1-\alpha\gamma) \} \geq 1-\alpha\gamma. 
\end{align*}
Then by the same arguments in \eqref{eq:propfinal}, the conclusion holds.

\newpage
\section*{Appendix C. Discussion on uniform asymptotic validity}\label{sec:appendixC}
In this section, we define a parameter space in order to define uniform asymptotic validity in the context of SPA tests, following \cite{andrews2010inference}. Then we provide an example showing that the modified hybrid test proposed in Definition \ref{def:modifiedhybridtest} does not enjoy uniform asymptotic validity. 

\subsection*{C.1. Parameterization for uniform asymptotic validity}
Uniform asymptotic validity is a stronger condition than pointwise asymptotic validity that we introduced in \hyperref[sec:setup]{Section 2}. Specifically, given a class of data generating processes or parameters, $\Theta$, and data $\{d_t\}^n_{t=1}$, a test $\phi_n=\phi_n(d_1, \ldots, d_n; \theta)$ for the null hypothesis $H_0$ is said to be uniformly asymptotically of level $\alpha$ if it satisfies
\begin{align*}
\limsup _{n \rightarrow \infty} \sup_{\theta\in \Theta_0} E\left[\phi_n\right] \leq \alpha
\end{align*}
where $\Theta_0$ is the set of parameters in $\Theta$ which satisfies the null hypothesis $H_0.$ Uniform asymptotic validity then implies that for any small number $\varepsilon>0$, the size of the test $\sup_{\theta\in \Theta_0} E\left[\phi_n\right]$ in the finite sample is less than $\alpha+ \varepsilon$ for sufficiently large sample size $n.$ In other words, the probability of committing type I error can be controlled by $\alpha+ \varepsilon$ uniformly over the data generating processes in the parameter space $\Theta_0$ for sufficiently large $n.$ This contrasts to the fact that pointwise asymptotic validity only provides asymptotic type I error control for a fixed data generating process and hence each data generating process or parameter $\theta$ requires different sample size $n$ in order to satisfy $E[\phi_n] \leq \alpha+ \varepsilon.$ 

For this reason, the importance of uniform asymptotic validity has been addressed in the literature of testing moment inequality where i.i.d. data are common. For further detail, refer to \cite{canayshaikh}. However, uniform asymptotic validity for time series data has not been yet fully discussed especially in the literature of SPA tests. We borrow the parameterization for dependent data from \cite{andrews2010inference} and show that the modified hybrid test does not enjoy uniform asymptotic validity. Whether there exists a reasonable parameter space which makes the modified hybrid test uniformly asymptotically valid or developing a general setup for SPA tests for uniform asymptotic validity could be interesting. Yet it is out of scope of this article, and we leave it for future research.

We restrict the parameter space $\mathcal{F}^{pt}$ which we use to prove pointwise asymptotic validity of the modified hybrid test in Proposition \ref{prop:modifiedtest} to $\mathcal{F}^{uf}$ following the parameterization in Section A.2 in \cite{andrews2010inference}. To do so, define $\theta = (\theta_1, \theta_2, \theta_3)$ as
\begin{align*}
    \theta_1   &\equiv D^{-1/2} E[d_t] \in \mathbb{R}^{M} \text{ with }D=diag(\Sigma)\\
    \theta_{2} &\equiv \text{vech}(\Omega_0) \in \mathbb{R}^{M(M-1)/2} \text{ where }\Omega_0 =D^{-1/2}\Sigma D^{-1/2}\\
    \theta_3 &\equiv F
\end{align*}
where the half-vectorization $\text{vech}(A)$ of a symmetric $M\times M$ matrix $A$ is the $M(M + 1)/2$-dimensional column vector obtained by vectorizing only the lower triangular part of $A.$ Note that there is one-to-one correspondence between $(\Sigma, F)$ and $\theta$ because $D=diag(\Sigma)$ can be recovered from $\theta_1$ and $\theta_3$ after extracting $E[d_t]$ from $F$, meaning that $\mathcal{F}^{pt}$ can be presented as
        \begin{align*}
            \mathcal{F}^{pt} = \{\theta: \text{Assumption \ref{as:stationarity}, \ref{as:weakconv}, \ref{as:bconsistency}, \ref{as:psd_cov}, and \ref{as:consistent_Sigma} are satisfied.}\}.
        \end{align*}
        
Consider a sequence $\{\theta_{n,h}\}^{\infty}_{n=1}$ such that 
\begin{align}
    \sqrt{n}\theta_{n, h, 1} \to h_1, \quad
    \theta_{n,h,2}=\text{vech}(\Omega_{n,h}) \to h_2, \quad
    \theta_{n,h,3}=F_{n,h} \label{eq:seq_theta}
\end{align}
for some $h=(h_1, h_2) \in (\mathbb{R}\cup \{\infty, -\infty\})^{M} \times \mathbb{R}^{M(M-1)/2}$. Note that $E[d_t]$ depends on the sample size $n$ because the distribution $F_{n,h}$ from which the sample $\{d_t: t=1,\ldots, n\}$ is drawn changes as $n$ varies, and thus from now on, we make this dependence explicit by adding subscript $\theta_{n,h}$ to expectations. Given $\theta_{n,h}$, let $\sigma_{n,h} \in \mathbb{R}^{M}$ with $\sigma_{n,h,m} = E_{\theta_{n,h}}[d_{t,m}]/\theta_{n,h,1}$. Consider the following conditions: under any $\{\theta_{n,h}\}^{\infty}_{n=1}$, 
        \begin{itemize}
            \item[(i)] $A_n = (A_{n,1},\ldots, A_{n,M})^t \overset{d}{\to} Z_{h_{2}}\sim N(0, \Omega_{h_2})$ as $n\to \infty$ where
            \begin{align*}
                A_{n,m} = \sqrt{n} \left(\hat{d}_{n,m} - \frac{1}{n} \sum^{n}_{t=1} E_{\theta_{n,h}}[d_{t,m}]\right) / \sigma_{n,h, m}\text{ for }m=1, \ldots, M
            \end{align*}
            and $\Omega_{h_2}$ is the correlation matrix defined by $h_2$;
            
            \item[(ii)] $\hat{\sigma}_{n,m} / \sigma_{{n,h}, m} \overset{p}{\to} 1$ as $n \to \infty$ for $m=1, \ldots, M$;
            
            \item[(iii)] $\hat{D}^{-1/2}_n \hat{\Sigma}_n \hat{D}^{-1/2}_n  \overset{p}{\to} \Omega_{h_2}$ as $n \to \infty$;
            
            \item[(iv)] conditions (i)-(iii) hold for all subsequences $\{w_n\}$ in place of $\{n\}$.
        \end{itemize}
Condition (i) guarantee that CLT-type of convergence result holds along the sequence $\{\theta_{n,h}\}^{\infty}_{n=1}$ and conditions (ii) and (iii) ensures consistency of the variance-covariance matrix estimator. 

Given the conditions above, we define $ \mathcal{F}^{uf} $ as
        \begin{align*}
             \mathcal{F}^{uf} \equiv \{\theta \in  \mathcal{F}^{pt}:\text{conditions (i)-(iv) hold for any }\{\theta_{n,h}\}^{\infty}_{n=1} \text{ defined in \eqref{eq:seq_theta}}.\}.
        \end{align*}
Furthermore, let $\mathcal{F}^{uf}_0$ be the subset of $ \mathcal{F}^{uf} $ which satisfies the null hypothesis $H_0$ in \eqref{eq:hypotheses}.

\subsection*{C.2. An counterexample}
We consider a simple example which shows the modified hybrid test in Definition \ref{def:modifiedhybridtest} is not uniformly asymptotically of level $\alpha$ under the paramterization $\mathcal{F}^{uf}.$

Let $M=2$ and $\gamma=0.5$. Consider a sequence of data generating processes $\{\theta_{n,h}\}^{\infty}_{n=1}$ in $\mathcal{F}^{uf}_0$. To make the dependence on $n$ explicit, we add subscript $n$ to the data. That is, for each $n\in \mathbb{N}$, $\{d_{n,t}\}^n_{t=1}$ is drawn from distribution $\theta_{n,h,3} = F_{n,h}$. Consider $\{\theta_{n,h}\}^{\infty}_{n=1}$ which satisfies two conditions. First, let $\theta_{n,h,3} = F_{n,h}$ satisfy
\begin{align*}
     E_{\theta_{n,h}}[d_{n,1}] = 0 \quad \text{ and }   E_{\theta_{n,h}}[d_{n,2}] = \frac{\tilde{h}}{\sqrt{n}} \text{ with some }\tilde{h}<0.
\end{align*}
Second, let $\Omega_{n,h}=I_2$ for all $n\in \mathbb{N}$ for simplicity. Then $h_2 = \text{vech}(I_2).$

To prove that the modified test is not uniformly asymptotically valid, we start by obtaining the asymptotic distribution of the test statistics. Conditions (i)-(iii) guarantee that
\begin{align*}
\sqrt{n}\hat{D}^{-1/2}_n( 
\hat{d}_{n,1} -   E_{\theta_{n,h}}[d_{n,1}],
\hat{d}_{n,2}-  E_{\theta_{n,h}}[d_{n,2}]
)^t \overset{d}{\to} Z\equiv \left(Z_1, Z_2\right)^t \sim N(0, I_2)
\end{align*}
along $\{\theta_{n,h}\}^{\infty}_{n=1}$ as $n\to \infty$, which gives
\begin{align*}
    \sqrt{n}\hat{D}^{-1/2}_n( 
\hat{d}_{n,1},
\hat{d}_{n,2}
)^t \overset{d}{\to} \left(Z_1, Z_2+\tilde{h}\right)^t \sim N( (0, \tilde{h})^t, I_2).
\end{align*}
Then the continuity of $S^r$ and $S^s$ defined in \eqref{eq:bootstat} and the continuous mapping theorem give the asymptotic distribution of the test statistics:
\begin{align*}
    \left(
    \begin{matrix}
     \tilde{T}^r_n\\
     \tilde{T}^s_n
    \end{matrix}
    \right) \overset{d}{\to} 
    \left(
    \begin{matrix}
    L^r\\L^s
    \end{matrix}
    \right) \equiv
    \left( 
    \begin{matrix}
    \max ( Z_1 \vee 0, (Z_2 +\tilde{h}) \vee 0 )\\
    \min(\max ( Z_1 \vee 0, (Z_2 +\tilde{h}) \vee 0 ), ~
    \max ( -Z_1 \vee 0, (-Z_2 -\tilde{h}) \vee 0 ))
    \end{matrix}
    \right) 
\end{align*}
as $n\to\infty.$

Next, we derive the probability limit of the critical values. Following the similar logic used in the proof of Proposition \ref{prop:modifiedtest}, we can show that the moment selecting vector $\hat{\psi}_{n, m}$ defined in \eqref{eq:momentselectingvec} is asymptotically negligible in that $\hat{\psi}_{n, m} \overset{p}{\to} 0$, and also that that for $q \in \{r,s\}$
\begin{align*}
\left| P^{\#} \{ S^q (\hat{\Omega}^{1/2}_n Z^{\#} + \hat{\psi}_{n}) \leq x\} - P \{ S^q (\Omega^{1/2}_0 Z^{\#}) \leq x\}\right| \overset{p}{\to} 0
\end{align*}
for any $x>0$ as $n\to \infty$ where $Z^{\#} \sim N(0_2, I_2)$ is independent from the sample and $P^{\#}$ denotes a probability measure conditioned on $(\hat{d}_n, \hat{\Omega}_n)$. Let $c^q(1-\alpha)$ be the $1-\alpha$ quantile of $P \{ S^q (\Omega^{1/2}_0 Z^{\#}) \leq x\}$ for $q\in \{r,s\}$. Note that
\begin{align*}
\left( 
\begin{matrix}
S^r(\Omega^{1/2}_0 Z^{\#})\\
S^s(\Omega^{1/2}_0 Z^{\#})
\end{matrix}
\right) 
\overset{d}{=}
\left(
    \begin{matrix}
    \max ( Z_1 \vee 0, Z_2 \vee 0 )\\
    \min(\max ( Z_1 \vee 0, Z_2  \vee 0 ), ~
    \max ( -Z_1 \vee 0, -Z_2 \vee 0 ))
    \end{matrix}
    \right).
\end{align*}
For $\alpha \in (0, 1/2) $ and for $ c^q(1-\alpha)>0 $, we have $\tilde{c}^q_n (1-\alpha)\overset{p}{\to} c^q(1-\alpha)$ again by the same argument used in the proof of Proposition \ref{prop:modifiedtest}.  

Comparing $ L^s$ and $S^s(\Omega^{1/2}_0 Z^{\#})$ tells us that we encounter the same problem as in Example 3.1: the distribution of $S^s(\Omega^{1/2}_0 Z^{\#})$ from which the probability limit of the critical value $c^s_n(1-\alpha)$ is obtained does not stochastically dominate the asymptotic distribution of $\tilde{T}^s_n$. This leads to uniform asymptotic invalidity. To see this, note that the rejection probability in this example converges as follows,
\begin{align*}
    E_{\theta_{n,h}}[\tilde{\phi}_n] \to P\{L^r >c^r(1-\alpha/2) \text{ or } L^s >c^s(1-\alpha/2)\} \text{ as }n\to\infty
\end{align*}
by the Slutsky theorem. Given the setting, we can obtain the value of this asymptotic rejection probability by numerical approximation. First, the probability limit of the critical values are: $c^r(1-\alpha/2) \approx 2.239$ and $c^s(1-\alpha/2)\approx 1.2171.$ Table C.1 presents how the limiting rejection probabilities vary depending on the value of $\tilde{h}$ when $\alpha=0.05$. The result is obtained by Monte Carlo simulations based on 50,000 repetitions and `Rejection Probability' denotes $P\{L^r >c^r(1-\alpha/2) \text{ or } L^s >c^s(1-\alpha/2)\}$. Clearly, the probability exceeds the significance level $\alpha$ and this phenomenon becomes more pronounced as $\tilde{h}$ decreases. Since the sequence of data generating processes $\{\theta_n\}^{\infty}_{n=1}$ from $\mathcal{F}^{uf}_0$ produces the limiting rejection probability $P\{L^r >c^r(1-\alpha/2) \text{ or } L^s >c^s(1-\alpha/2)\}$ of which values are larger than the significance level $\alpha$ at 0.05, we conclude that the modified hybrid test is not uniformly asymptotically of level $\alpha $ under the parameterization $\mathcal{F}^{uf}.$

\setcounter{table}{0}
\renewcommand{\thetable}{C.\arabic{table}} 

\begin{table}[htbp]\label{tab:appendixCtab}
\begin{center}
    \caption{Simulated Probabilities}
    \begin{tabular}{cccccc}\hline\hline
    $\tilde{h}$     & -5    & -4    & -3    & -2    & -1\\\hline 
    Rejection Probability & 0.1118 & 0.1115 & 0.1084 & 0.0914 & 0.057 \\\hline
    \end{tabular}%
\end{center}
\end{table}%

\end{document}